\documentclass{aa}  

\usepackage{graphicx}
\usepackage{txfonts}
\usepackage{natbib}
\usepackage{amsmath} 
\usepackage{multirow}
\usepackage{booktabs}
\usepackage{float}
\usepackage{hyperref} 
\usepackage{graphics} 
\usepackage{subfig} 
\usepackage{xcolor}
\usepackage{enumitem}
\usepackage{amsmath}
\usepackage{bm}
\usepackage{mathtools}
\usepackage{soul}
\usepackage{lscape}
\usepackage{rotating}
\usepackage{caption}

\usepackage{adjustbox}

\begin{document} 

   \title{Planet Earth in reflected and polarized light}

   \subtitle{III. Modeling and analysis of a decade-long catalog of Earthshine observations}

   \author{Giulia Roccetti
          \inst{1,2}
          \and
          Michael F. Sterzik
          \inst{1}
          \and
          Claudia Emde
          \inst{2,3}
          \and
          Mihail Manev
          \inst{2}
          \and
          Stefano Bagnulo
          \inst{4}
          \and
          Julia V. Seidel
          \inst{5,6}
          }

   \institute{
    European Southern Observatory, Karl-Schwarzschild-Straße 2, 85748, Garching near Munich, Germany\\
    \email{giulia.roccetti@eso.org}
    \and
    Meteorologisches Institut, Ludwig-Maximilians-Universität München, Munich, Germany
    \and 
    Rayference, Rue d’Alost 7, 1000, Bruxelles, Belgium
    \and
    Armagh Observatory and Planetarium, College Hill, Armagh BT61 9DG, Northern Ireland, UK
    \and
    European Southern Observatory, Santiago, Chile
    \and
    Laboratoire Lagrange, Observatoire de la Côte d’Azur, CNRS, Universit\'e Côte d’Azur, Nice, France}
    
   \date{Received 31 May 2025; accepted 24 July 2025}

  \abstract{Earthshine observations offer a unique opportunity to study Earth as an exoplanet seen from the Moon. As the Sun-Earth-Moon geometry changes, Earth can be observed as a spatially unresolved exoplanet at different phase angles, providing important context for future observations of Earth-like exoplanets. Here, we present a catalog of Earthshine polarization spectra obtained with FORS2 on the VLT, covering diverse scenes, surface conditions, cloud properties, and weather patterns for over a decade. For the first time, we model this extensive dataset in detail using a homogeneous modeling framework. Previous efforts to model some of these spectra struggled to reproduce the observed polarization continuum, even with advanced 3D radiative transfer models incorporating satellite-derived surface and atmospheric data. We improve upon this with a state-of-the-art 3D model that includes subgrid cloud variability, wavelength-dependent surface albedo maps, and an accurate treatment of ocean glint. Our simulations successfully reproduce most observed spectra to a much higher precision than previously possible. Additionally, our statistical analysis reveals that the spectral slope in the visible can distinguish between ocean and mixed surfaces in both reflected and polarized light, which is not possible using broadband filters alone. Polarized light at large phase angles, beyond the Rayleigh scattering regime, is particularly effective in differentiating oceans from land, unlike reflected light. While the vegetation red edge is more pronounced in reflectance, it remains detectable in polarization. We also identify correlations between cloud optical thickness and the polarized spectral slope, and between cloud cover and broadband B–R differences in reflected light, demonstrating the diagnostic power of these observations. This catalog and its modeling highlight the potential of polarization for characterizing Earth-like exoplanets. From polarization alone, we can infer oceans, vegetation, and an active water cycle, key indicators of a habitable planet.
  }


   \maketitle

\section{Introduction}
\label{sec:intro}

The search for small, habitable exoplanets is a primary objective of upcoming observatories such as the Extremely Large Telescope (ELT) and the proposed mission concept Habitable Worlds Observatory (HWO). To fully interpret future data from these missions, we must significantly improve our ability to model observations. Earth remains the only known example of a habitable planet, and it provides a unique benchmark to address a fundamental question: how unique is our planet? While not all habitable exoplanets will closely resemble Earth, many are expected to share key features, such as oceans, clouds, and surface heterogeneities. Consequently, studying Earth as an exoplanet is vital for developing observational strategies and interpretative tools. A range of techniques have been used to examine Earth’s disk-integrated properties, offering valuable insight for future photometric and spectroscopic observations of Earth-like exoplanets \citep{Robinson2018}.\\
\indent The first effort to observe Earth as a planetary body using a space-based platform was carried out during the Galileo spacecraft’s flybys of Earth \citep{Johnson1992}. Spatially resolved imagery and spectra revealed important atmospheric and surface characteristics. In particular, \cite{Sagan1993} identified specular reflection indicative of surface liquid water, high column densities of O$_2$ in disequilibrium with CH$_4$, and the vegetation red edge (VRE), a sharp increase in reflectivity beyond 700~nm associated with chlorophyll. These findings pointed to a biologically active planet with surface diversity and an atmosphere influenced by life.\\
\indent When observing exoplanets, however, we cannot resolve them spatially. Instead, we receive light from the entire planet averaged into a single pixel. In such cases, the viewing geometry plays a critical role in shaping the reflected light signal. For instance, regions near the terminator contribute less to the disk-integrated signal, and features such as specular reflection depend on the directional reflectance of the surface. Properly accounting for geometry is essential to extracting accurate information from reflected light observations, a crucial aspect of the ELT.\\
\indent Since the mid-20th century, Earth has been continuously monitored by a growing number of satellites. By stitching together high-resolution satellite imagery, it is possible to build datasets of disk-integrated Earth observations \citep{Hearty2009}, offering broad spectral, temporal, and spatial coverage. However, this approach does not replicate the conditions of typical exoplanet observations due to temporal gaps influenced by varying weather patterns and constraints in viewing geometry, especially for low Earth orbit satellites.\\
\indent The ideal approach to mimic direct observations of the Earth as an exoplanet would be to acquire photometry and spectroscopic measurements for a truly distant Earth, from distances beyond low-Earth orbit or geostationary satellites.
For example, the EPOXI mission captured disk-integrated Earth spectra, revealing key spectral features and enabling the mapping of unresolved planetary surfaces \citep{Robinson2011}. Observations from NASA’s LCROSS mission helped quantify the effects of ocean glint and ozone absorption on Earth’s disk-integrated reflectance spectra at various phase angles (i.e., the star-planet-observer angle) \citep{Robinson2014}. Additionally, NASA's DSCOVR mission, located at the Earth-Sun L1 Lagrange point, provides continuous imaging of Earth’s fully illuminated hemisphere that can be used to obtain disk-integrated reflected light spectra \citep{Kofman2024}. However, these measurements lack the phase angle diversity needed to replicate typical exoplanet observation conditions. Furthermore, most spacecraft datasets focus on photometry rather than spectroscopy, and disk-integrated observations beyond quadrature remain limited, with LCROSS being a rare exception \citep{Robinson2018}.\\
\indent An alternative method to study Earth as an exoplanet is through Earthshine, which is sunlight scattered by Earth’s dayside and reflected off the darker portion of the visible Moon. The Moon acts as a diffuse reflector, integrating light from the illuminated hemisphere of Earth. As early as the Renaissance, Leonardo Da Vinci noted that the faint glow of the darker portion of the visible Moon on clear nights was due to light reflected from Earth \citep{DaVinci1510}, and named it Earthshine. Later, Galileo Galilei was the first to observe Earthshine with a telescope and to explain that different regions of Earth reflected light differently, emphasizing the role of planetary albedo \citep{Galilei1632}. The first quantitative measurements of Earthshine were carried out by \citet{Dubois1947}, who presented Earth’s phase curve and demonstrated that broadband reflectivity varies significantly with phase angle due to cloud variability.\\
\indent Modern Earthshine studies have progressed from photometry to spectroscopy, allowing for detailed characterization of Earth’s disk-integrated spectrum \citep{Goode2001}. These studies have revealed daily to decadal variability in Earth’s reflectivity \citep{Palle2003, Palle2009, Palle2016}, including changes in the VRE linked to cloud cover and surface vegetation \citep{Arnold2002, Seager2005, Montanes-Rodriguez2006}. Spectral features associated with habitability and life, such as H$_2$O, O$_2$, and CH$_4$, have been identified in Earthshine spectra spanning 0.7–2.4~$\mu$m \citep{Turnbull2006}.\\
\indent Despite its advantages, Earthshine observations are not without challenges. The Moon is not a perfect Lambertian reflector, and the signal must pass through Earth’s atmosphere twice, introducing telluric contamination. One way to mitigate this is through Earthshine observations using spectropolarimetry, thus obtaining the fractional polarization of Earthshine as a function of wavelength. \cite{Sterzik2012} obtained the first such observations that compare two different Earth's scenes: one with the illuminated side over the Pacific Ocean, and the other featuring a mixture of land and ocean surfaces. With these observations, changes in Earth’s cloudiness, the presence of an ocean, and even the VRE, a biosignature caused by chlorophyll, were detected. Independent measurements of Earthshine obtained from different geographical regions of the world were conducted by \cite{Takahashi2013} and \cite{Bazzon2013}, showing overall good agreement with the initial polarization spectra presented in \cite{Sterzik2012}. \cite{Miles-Paez2014} further extended the wavelength coverage of Earthshine polarimetric measurements into the near-infrared (NIR), demonstrating sensitivity to molecular absorption lines. \cite{Sterzik2019} expanded the phase angle coverage of Earthshine observations, presenting 33 spectra obtained during an observational campaign and constructing polarized phase curves of Earth as an exoplanet in the visible (VIS) and NIR. For Earthshine, the phase angle changes with the relative positions of the Sun, Earth, and Moon. As a final extension to phase angle coverage, \cite{Sterzik2020} observed Earthshine at small phase angles (around 30°–40°) to probe the cloudbow feature, which depends on the properties of cloud droplets. This led to the first detection of cloudbow features in the disk-integrated Earth. \cite{Sterzik2020} used this feature to retrieve the refractive index and sizes of cloud droplets on Earth. This approach was proven highly successful on Venus, where disk-integrated observations at small phase angles revealed that the planet's thick clouds are composed of sulfuric acid with particle sizes of about 2~$\mu$m \citep{Hansen&Travis1974}. Moreover, \cite{Takahashi2021} have further extended the phase coverage of Earthshine observations.\\
\indent In parallel with observational advances, modeling efforts have also progressed. Early models by \citet{Stam2008} considered horizontally homogeneous planets in plane-parallel atmospheres. These were extended by \citet{Karalidi2012} to include surface and cloud heterogeneities, with further work exploring the cloudbow as a diagnostic of liquid water in Earth-like atmospheres \citep{Karalidi2012c}. Monte Carlo radiative transfer models capable of simulating both intensity and polarization were developed by \citet{Garcia-Munoz2015} and \citet{Emde2017}, who demonstrated the sensitivity of polarization to clouds, aerosols, and surface properties. Additionally, \cite{Emde2017} demonstrated that light reflected by ocean surfaces in the sunglint region causes high degrees of polarization, consistent with the findings of \cite{Sterzik2012}. Recent models by \citet{Trees2022} predicted ocean glint signatures in polarization, including distinct absorption dips across the 950~nm water band. However, \citet{Gordon2023} found that matching observed polarization spectra from \cite{Miles-Paez2014} remains challenging, in part due to oversimplified surface models that neglect the ocean glint. Their follow-up work \citep{Gordon2025} explored Earth’s polarized appearance across geological epochs, showing that polarization offers greater discriminative power for cloud and haze properties than intensity-only observations.\\
\indent In this paper, we present the first extensive modeling effort of a large catalog of Earthshine observations in polarization. This constitutes a critical step toward benchmarking model performance and understanding what Earth would look like as an unresolved exoplanet. Our modeling framework, based on \cite{Roccetti2025a}, allowed us to simulate each Earthshine spectrum using cloud and surface data from the exact time of the observations. We also improve upon the data reduction and analysis techniques from \cite{Sterzik2019, Sterzik2020}, creating a higher-quality, more uniform dataset. Our focus is on the spectral continuum and on features such as the ocean glint, the VRE, and cloud properties. Finally, we assessed the diagnostic potential of polarization relative to intensity-only observations and explored what could be learned about Earth when seen as a distant, spatially unresolved planet.

\section{Earthshine observations}
\label{sec:observations}

Our catalog of Earthshine polarization observations consists of 53 spectra covering phase angles from 37$^\circ$ to 136$^\circ$, previously published in \cite{Sterzik2019} and \cite{Sterzik2020}. All observations were obtained using the FORS2 instrument \citep{Appenzeller1998}, a low-resolution spectrograph with polarimetric optics, mounted on the Antu telescope at the ESO Very Large Telescope (VLT) at Cerro Paranal, Chile. Table \ref{table} reports the date, phase angle, and observational configurations of all Earthshine observations from \cite{Sterzik2012}, \cite{Sterzik2019}, and \cite{Sterzik2020}.\\
\indent FORS2 is equipped with Wollaston prisms and a rotating retarder waveplate, allowing the measurement of the wavelength-dependent reduced Stokes parameters $P_Q = Q/I$ and $P_U = U/I$, from which the total fractional linear polarization is calculated as:
\begin{equation}
    P = \sqrt{P_Q^2 + P_U^2}.
\end{equation}
The spectra were collected using two different grisms:
\begin{itemize}
    \item The 300V grism was used for 45 spectra, covering the 420-920~nm range with a spectral resolution of $\approx 220$ using a 2\arcsec slit.
    \item The 600I grism was used for the remaining 8 spectra, covering 670-930~nm with a spectral resolution of 750.
\end{itemize}
Observations targeted the darker portion of the visible Moon, with the FORS2 detector oriented east-west along the lunar limb. The first detector chip contains five 22$\arcsec$ long slitlets positioned across the lunar surface. A 4$\arcsec$ gap separates it from the second chip, which contains four slitlets pointed at the empty sky for background subtraction. The lunar limb was consistently positioned in the gap between the two detector chips.\\
\indent Polarimetric data were acquired using the beam-swapping technique \citep{Bagnulo2009}, which reduces instrumental systematics typical of dual-beam polarimetry. This was achieved by acquiring exposures at 16 retarder waveplate angles, from 0$^\circ$ to 337.5$^\circ$ in 22.5$^\circ$ increments.\\
\indent Additionally, we applied a correction for lunar depolarization. The lunar depolarization factor $\epsilon(\lambda)$ is defined as:
\begin{equation}
    \epsilon(\lambda) = \frac{P_{\text{out}}(\lambda)}{P_{\text{in}}(\lambda)},
\end{equation}
where $P_{\text{in}}(\lambda)$ is the fractional polarization of light incident on the Moon, and $P_{\text{out}}(\lambda)$ is the polarization of the reflected Earthshine.\\
\indent Following \cite{Bazzon2013}, we compute $\epsilon$ using the lunar albedo at the observation site, as a function of wavelength. The polarization efficiency $\log \epsilon$ depends on both the lunar albedo at 603~nm ($\log a_{603}$) and the wavelength ($\log \lambda$), and is given by:
\begin{equation}
    \log \epsilon (\lambda, a_{603}) = -0.61 \log a_{603} - 0.291 \log \lambda\, [\mu \mathrm{m}] - 0.955.
\end{equation}
As in \cite{Sterzik2019} and \cite{Sterzik2020}, we adopted this method and used their $a_{603}$ values, extracted by comparison with the lunar albedo maps of \cite{Velikodsky2011}.\\
\indent For details on data acquisition and reduction, including preprocessing, flat-fielding, and background subtraction, we refer to \cite{Sterzik2019}. In particular, flat-fielding is essential for Earthshine spectropolarimetry since the sky background must be interpolated on chip 2 and linearly extrapolated to chip 1 to be subtracted from the Earthshine signal. This step relies on the assumption that Moonshine intensity decreases linearly with distance from the terminator. However, in practice, we observed an anti-correlation between intensity and fractional polarization along the slit. To address this, we developed a new procedure to average the five slitlets on chip 1 and improve the quality of the resulting polarization spectra. To do this, we searched for the best functional form to fit the spectra (and each slitlet), excluding the absorption lines. Looking at the spectra in $\log(P)$–$\log(\lambda)$ space, we determined that the spectra follow two power laws with a kink shifting its wavelength for each epoch. We then derived the following functional form:
\begin{equation}
    P(\lambda) = P_1 \left( \frac{\lambda}{\lambda_1} \right)^{\gamma} \cdot (1 - s(\lambda)) + P_2 \left( \frac{\lambda}{\lambda_2} \right)^{\beta} \cdot s(\lambda),
    \label{eq:fit} 
\end{equation}
where $\gamma$ and $\beta$ are the fitted slopes in the VIS and NIR, $\lambda$ is the wavelength, $\lambda_1$ = 600~nm and $\lambda_2$ = 800~nm are fixed to facilitate the fit and $P_1$ and $P_2$ are left as free parameters. The $s(\lambda)$ function is a sigmoid defined as
\begin{equation}
    s(\lambda) = \frac{1}{1 + e^{-k(\lambda - \lambda_0)}},
\end{equation}
with $\lambda_0$ representing the position of the kink and $k$ representing the strength of the transition. Both $\lambda_0$ and $k$ are also free parameters of the fit. These two power laws weighted by a sigmoid function ensure the fitting of all spectra, together with extracting their spectral slopes.\\
\indent The improved slitlet-averaging procedure follows these steps:
\begin{enumerate}
    \item We cut the spectra to the 450-900~nm range for the 300V grism and 680--920~nm for the 600I grism. Each slitlet is then corrected for lunar depolarization, as explained below.
    \item Slitlets showing unphysical behavior, such as negative polarization values, extreme red-end increases, or spurious bumps, are removed. This is particularly necessary for the H and G epochs, which correspond to the smallest and largest phase angles, respectively (see Table \ref{table}), and are the most problematic.
    \item We calculated the standard deviation across the remaining slitlets (typically five), averaged over all wavelengths. If the standard deviation was below 1\%, the data are deemed high quality, and we computed the average over all slitlets. This is the case for most spectra at phase angles smaller than quadrature (e.g., epochs E, I, J, K).
    \item For the remaining spectra, we selected the slitlet with the lowest noise, computed as the random mean squared error (RMSE) compared to the fit, as reference (best slitlet). We then averaged only those slitlets that lie within 1$\sigma$ of the best slitlet in VIS slope ($\gamma$), NIR slope ($\beta$), and average polarization distance. This excluded slitlets with inconsistent spectral shapes or anomalous polarization levels, which may indicate contamination. Typically, the last slitlet (closest to the limb) was found to be the least noisy and it was assumed as the best slitlet. At high phase angles (e.g., G epochs), this process often resulted in only the fifth slitlet being retained.
\end{enumerate}
Earthshine observations become increasingly challenging with larger phase angles, particularly beyond quadrature, as the Moon's dayside terminator approaches the observational field, increasing Moonshine contamination. Observations above $\alpha \approx 130^\circ$--$140^\circ$ from the ground become nearly unfeasible. However, these geometries are crucial to studying the ocean glint feature. Our catalog includes three such epochs (G.7, G.8, and G.9, all with $\alpha > 135^\circ$). Due to their particularly high noise and uncertainty, we treated them separately: for G.7, only slitlet 4 was used, while for G.8 and G.9, we averaged slitlets 3, 4, and 5.\\
\indent A similar issue arises at very small phase angles ($\alpha < 40^\circ$), when the thin crescent Moon is visible only briefly at twilight, resulting in short observing windows and high airmass. These conditions are also scientifically valuable, as they reveal the cloudbow feature, which provides key information on cloud microphysics. Epoch H.1, our lowest phase angle spectrum ($\alpha = 37^\circ$), was carefully analyzed. We averaged slitlets 2 and 5 to obtain a consistent result from components with different spectral slopes.\\
\indent In contrast, we excluded the J.1 and K.1 epochs from our updated catalog. These spectra, previously identified as uncertain in \cite{Sterzik2020}, were acquired at very low phase angles (33$^\circ$ and 35$^\circ$, respectively) during brief twilight windows, which prevented the completion of all 16 retarder positions. Only two positions were recorded, resulting in significant systematic errors that rendered our slitlet averaging procedure inapplicable.

\section{3D radiative transfer simulations}
\label{sec:simulations}

For all observational epochs listed in Table \ref{table}, we performed 3D radiative transfer simulations using surface and atmospheric conditions corresponding precisely to the time of observation. To achieve this, we employed the Monte Carlo code MYSTIC (Monte Carlo code for the phYsically correct Tracing of photons in Cloudy atmospheres; \citep{Mayer2009}), which is part of the libRadtran software package \citep{Mayer&Kylling2005, Emde2016}. MYSTIC supports 3D Earth-like atmospheres, includes full Stokes vector calculations to account for polarization, and allows for 2D inhomogeneous surface representations.\\
\indent A modeling setup for using MYSTIC to simulate Earthshine observations was first introduced by \cite{Emde2017}. However, it has been substantially improved in a previous paper of this series \citep{Roccetti2025a}, which presented a more sophisticated framework for simulating both cloud and surface properties. In particular, cloud modeling now incorporates a 3D Cloud Generator algorithm that captures sub-grid cloud variability and inhomogeneities, using input from ECMWF ERA5 cloud data.\\
\indent Regarding surface albedo treatment, oceans are modeled as specular reflectors using a BPDF approach for reflected light and polarized light. Land surfaces are treated as Lambertian reflectors with albedo values varying with wavelength, based on the hyperspectral albedo maps presented in \cite{Roccetti2024}. These maps combine MODIS satellite data with a comprehensive set of in-situ and laboratory spectra of various soils and vegetation types, enabling wavelength-dependent albedo modeling, which is crucial for studying features such as the VRE. \\
\indent In \cite{Roccetti2025a}, we provide a detailed analysis of how the improved treatments of clouds and surface properties affect the spectra and phase curves of an ocean and an Earth-like exoplanet. In \cite{Roccetti2025b}, we evaluate how these more realistic and detailed models compare to simulations that use spatially averaged homogeneous conditions and simplified representations of cloud and surface features.\\
\indent For the simulations, the ERA5 cloud properties are rounded to the nearest full hour relative to the central time of each observation. The Sun and Moon coordinates used to replicate the viewing geometry are taken at the midpoint between the start and end times of each observational epoch listed in \cite{Sterzik2019} and \cite{Sterzik2020}, based on data from the Earth-Moon viewer \footnote{\url{https://www.fourmilab.ch/earthview/}}.

\section{Comparison between observations and simulations}
\label{sec:comparison}

\begin{figure*}
    \centering
    \includegraphics[width=1\linewidth]{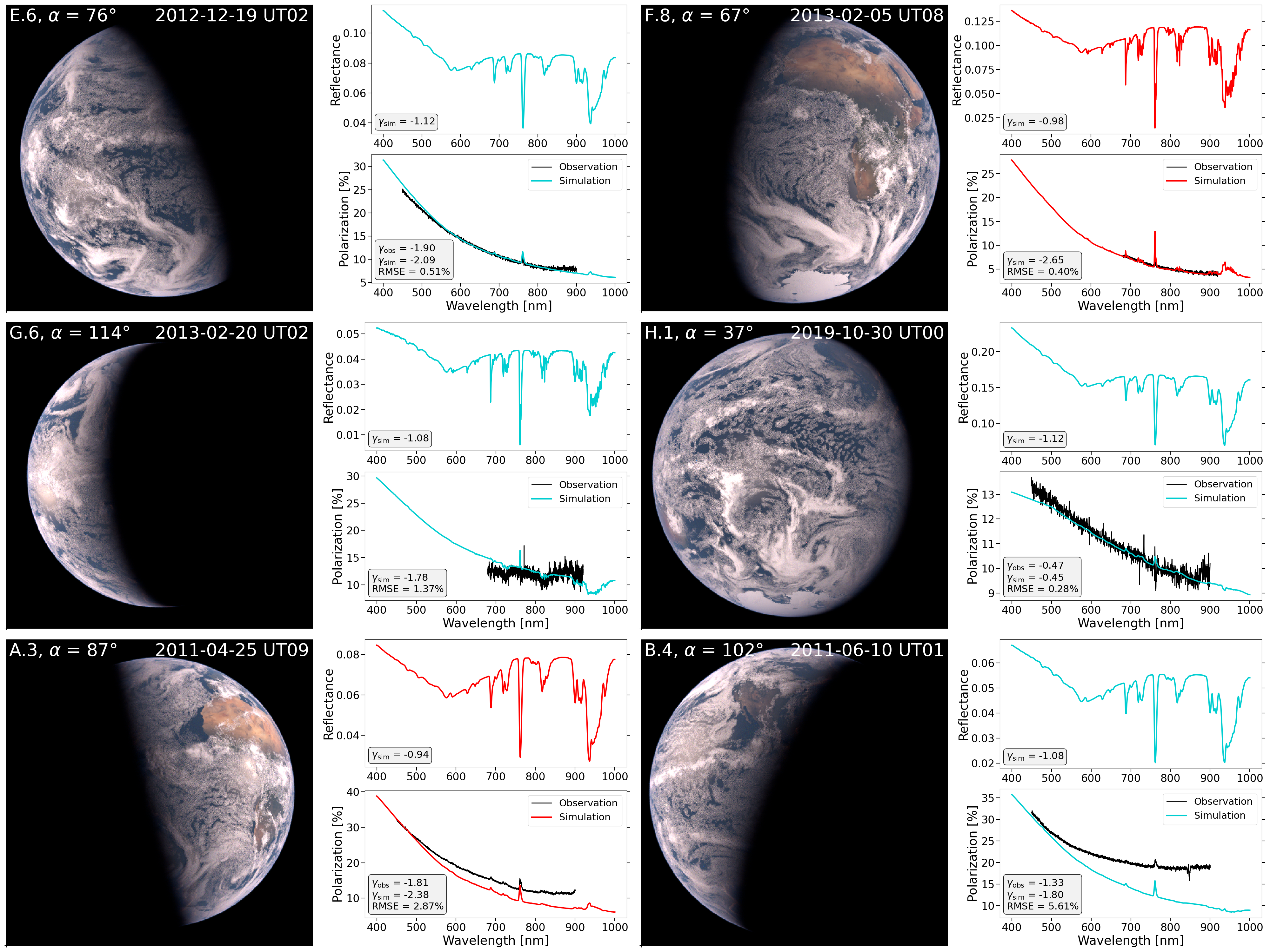}
    \caption{Selected examples of observed polarized Earthshine spectra, along with corresponding simulations of reflected and polarized light, as presented in Appendix \ref{appendix-B}. For each spectrum, we include a simulated observation image, fit the power-law slopes $\gamma$, and report the RMSE between simulations and observations.}
    \label{fig:comparison}
\end{figure*}

In Appendix \ref{table}, we present the comparison between the full catalog of Earthshine observations and their corresponding simulations. In addition to the spectra, we simulated an image of Earth for the same time and viewing geometry as the observation, using the actual cloud properties from ERA5 at the time of the observations, the Sun–Earth–Moon orientation, and wavelength-dependent albedo maps from HAMSTER on the specific observation date. The simulation of these images follows the method described in \cite{Roccetti2025a}. In addition to the image, we present the simulated reflected light spectra in units of reflectance, defined as $A_\text{g} \cdot g(\alpha)$, as well as a comparison between the simulated and observed polarized spectra, expressed in percent polarization. All comparisons are displayed in Figs. \ref{fig:collage1} to \ref{fig:collage5}. The simulations span the 400–1000 nm spectral range and are adjusted in resolution to match the specific setup used during each observation. The axes for reflectance and polarization are scaled to match the extrema of observations and simulations independently, in order to enhance the visual clarity of the comparisons.\\
\indent Each observational spectrum is color-coded to indicate its viewing geometry. Spectra shown in turquoise correspond to what we define as the Pacific configuration, as described in \cite{Sterzik2019}, which represents Earthshine observations taken at the beginning of the night from Paranal with the Pacific Ocean as the dominating scene and very little land visible, thus resembling the appearance of an ocean planet. Spectra shown in red correspond to the Atlantic configuration, which includes a mixture of land and ocean, with parts of the Atlantic Ocean, Africa, Europe, South America, or the Middle East in view.\\
\indent For each epoch, we also fit a double power-law function as described in Section \ref{sec:observations}, reporting the power-law exponent in the visible range (400–700~nm) for both the simulated reflected and polarized light, and for the observed polarized spectra taken with the 300V grism, as the 600I grism covers only wavelengths longer than 680~nm. Additionally, we calculate the RMSE of each observed and simulated polarization spectrum within the spectral range where data is available to assess how well the simulations reproduce the observations.\\
\indent In Fig. \ref{fig:comparison}, we present a representative selection of the catalog epochs to illustrate our sample. Throughout the paper, spectra shown in red correspond to an Atlantic scenario, while those in cyan correspond to a Pacific scenario. In the top left panel, we show epoch E.6, which is a Pacific scenario with nearly no land in view (only a small portion of Antarctica), observed at a phase angle $\alpha = 76^\circ$ with grism 300V, covering the 450–900 nm range (note that the spectral edges are cut; see Section \ref{sec:observations}). This epoch shows very strong agreement with the simulation, as is typical for Pacific (ocean planet) configurations at phase angles smaller than quadrature. In the top right panel, we show an Atlantic (mixed) scenario also at a phase angle smaller than quadrature, observed with grism 600I. The 600I grism does not include the shorter visible wavelengths, where Rayleigh scattering dominates, and has a higher spectral resolution than the 300V; thus, the simulation was adjusted in its spectral resolution. In the second row, we present two of the most challenging spectra to model: epoch G.6, with a very high phase angle, and epoch H.1, with the smallest phase angle in our sample. Both spectra are noisier than previous examples. This is due to increased contamination from Moonshine at phase angles beyond quadrature and to the challenging lunar position at small phase angles. Despite these challenges, both spectra show good agreement between observations and simulations in polarization, as confirmed by the RMSE and slope comparisons.\\
\indent In the final row, we present epochs A.3 and B.4, which were previously discussed in \cite{Sterzik2012}. These spectra, and all spectra from epochs A, B, and C, exhibit significantly flatter spectral slopes compared to observations obtained in later years. Even though we cannot reproduce the spectral slopes of A.1 to C.2, we achieve a clear improvement in matching the observations compared to earlier attempts using the models of \cite{Stam2008} and \cite{Emde2017}. This improvement arises from two key factors: the updated slitlet-averaging procedure, particularly important for phase angles beyond quadrature, and the improved modeling of cloud and surface properties as introduced in \cite{Roccetti2025a}.\\
\indent Previous modeling efforts by \cite{Emde2017} and \cite{Gordon2023} attempted to simulate Earthshine polarization spectra. The former used MYSTIC as we do, while the latter used the VSTAR and DAP models. Both focused on a limited number of observations: \cite{Emde2017} examined epochs A.3 and B.4 from \cite{Sterzik2012}, while \cite{Gordon2023} analyzed a single spectrum from \cite{Miles-Paez2014} extending into the NIR up to 2500 nm. In both cases, the models failed to match the observed slope in visible polarization. Moreover, \cite{Gordon2023} could not reproduce the polarization continuum, largely due to the treatment of the ocean as a dark Lambertian surface. However, the importance of the ocean glint effect had already been demonstrated by \cite{Emde2017}, who showed its strong impact in disk-integrated Earthshine polarization observations at large phase angles. In contrast, our simulations achieve an excellent match for all spectra from epochs E.1 to K.3, something never accomplished before. High phase angle observations, previously considered particularly difficult due to the influence of ocean glint, now show strong agreement with simulations. This success is largely due to the implementation of the 3D Cloud Generator developed by \cite{Roccetti2025a}, which captures cloud subgrid variability and heterogeneity beyond the already detailed ERA5 cloud data. For example, at a phase angle of $\alpha = 120^\circ$, the inclusion of the 3D Cloud Generator significantly impacts the polarization spectral slope and enhances the continuum, which allows us to match the observed Earthshine spectra.\\
\indent We are also now able to accurately simulate spectra from Atlantic (mixed) viewing geometries, which had not been successfully reproduced in the past. This is made possible by the combination of a BPDF for oceans and a Lambertian model for land surfaces, as well as by using wavelength-dependent albedo maps from HAMSTER \citep{Roccetti2024}. These maps account for realistic mixtures of soil and vegetation types and result in a significantly lower modeled albedo for forested and desert regions compared to those used in \cite{Gordon2023} and \cite{Kofman2024}. As shown in \cite{Roccetti2025a}, this refinement is essential for matching the observed continuum.\\
\indent Regarding the earlier observational epochs from A.1 to C.2, which we are unable to match, we explored several possible explanations. Notably, the B epochs, despite corresponding to Pacific (ocean) configurations, appear much flatter than later Pacific observations taken at similar phase angles (e.g., G.1 to G.4). A similar pattern is seen for the Atlantic (mixed) configurations in the A and C epochs, which differ significantly in both slope and polarization level from the F epochs at a similar phase angle. While these observations followed the same acquisition and reduction techniques, one notable difference is the use of screen flats for calibration instead of sky flats. However, \cite{Sterzik2019} found no significant difference when comparing screen and sky flat calibrations in the E epochs. Additionally, we are not aware of any major changes in the FORS2 instrument or its calibration procedures between 2011 and the following years. For an alternative explanation, we investigated whether any global atmospheric event could have affected the scattering properties of Earth’s atmosphere during this period, but we found no evidence of major volcanic eruptions. We also checked whether this flattening of the spectra could depend on the sky condition over Paranal. A strong El Ni\~no event occurred in 2011, which is known to influence observing conditions over Paranal, particularly in terms of turbulence, sunset temperature anomalies, and precipitable water vapor \citep{Seidel2023}. El Ni\~no’s effects can persist over time, potentially affecting observations up to epoch C.2 in October 2012. While 2019 was also an El Ni\~no year, its impact was weaker, and all observations from that year were made at small to moderate phase angles. Within epochs A.1 to C.2, a trend toward flatter spectra with increasing phase angle is also evident. This is consistent with independent observations obtained in 2011 by \cite{Takahashi2013} and \cite{Bazzon2013}, from Japan and Switzerland, respectively, who also reported a flattening of the spectra at large phase angles, as already discussed in \cite{Sterzik2019}.\\
\indent Because we have no definitive explanation for the behavior of these early spectra and find them problematic, we show them in the subsequent statistical analysis (Section \ref{sec:statistics}) for completeness, but exclude them from the actual calculations. In the next plots, they are displayed with transparent colors to clearly distinguish them from the rest of the sample used to determine disk-integrated properties of Earth.

\section{Population studies}
\label{sec:statistics}

Building on our extensive catalog of Earthshine observations and corresponding simulations in both reflectance and polarization, we now explore whether significant correlations can be identified between Earth's cloud and surface properties and the observed and simulated spectral features. To this end, we focus exclusively on the spectral continuum of the observations and test several diagnostic metrics: the spectral slope in the visible (VIS) range, a broadband color difference between the B and R filters (as defined by the typical Johnson filters), and the continuum value at a single reference wavelength. For the B filter we use the spectral range from 435 to 455~nm, while for the R filter from 645 to 665~nm.\\
\indent Furthermore, to evaluate the detectability of the VRE, we compute two vegetation indices: the Normalized Difference Vegetation Index (NDVI) \citep{Tinetti2006b} and the Polarized Difference Vegetation Index (PDVI) \citep{Sterzik2019}. In all subsequent plots, simulated data points are marked with stars, while observed polarization measurements are shown as dots. The different Earth viewing geometries are color-coded: Pacific (ocean-dominated) configurations in turquoise, and Atlantic (mixed land-ocean) configurations in red.\\ 
\indent Error bars on the observational data are omitted from the figures, as they are smaller than the size of the plotted data points. The uncertainty associated with individual slitlets accounts for both measurement errors and the contribution from lunar depolarization, and is propagated through the slitlet averaging procedure. Errors for each spectrum and calculated quantity are provided in the public database (see Sec. \ref{sec:data}). For the simulations, error bars are similarly small, within the thickness of the plotted points. A more detailed discussion of the simulation uncertainties is available in \cite{Roccetti2025a}.\\
\indent Although a comparison between observed and simulated reflectance spectra would be informative, well-calibrated reflectance data are not available for this Earthshine catalog. Nevertheless, as discussed in \cite{Roccetti2025a, Roccetti2025b}, polarization spectra and phase curves are more sensitive to surface and cloud properties. For example, the implementation of the 3D Cloud Generator (3D CG) affects the spectral slope in polarization but causes minimal changes in reflectance. In reflectance, the main effect is a shift in the continuum level, which requires high-precision calibration, unavailable in this dataset.

\subsection{Phase curves to distinguish between an ocean and mixed surface scenario}
\label{sec:ocean-land}

\begin{figure*}
    \centering
    \includegraphics[width=0.95\linewidth]{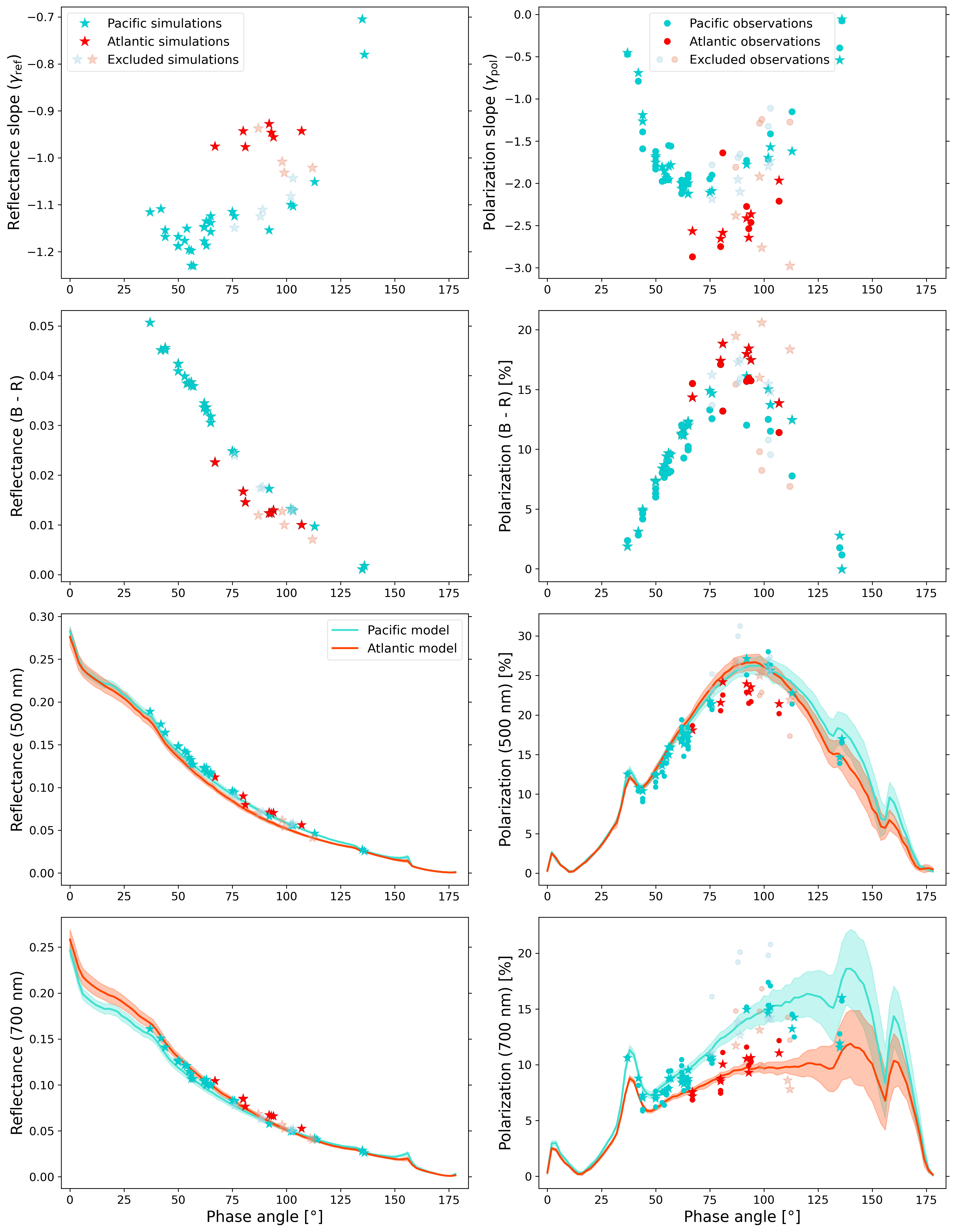}
    \caption{Reflectance simulations (first column) and polarized observations and simulations (second column) from our Earthshine catalog. We examine different diagnostic metrics: the spectral slope in the visible range (first row), the B–R broadband difference (second row), and the continuum reflectance or polarization at single wavelengths ($\lambda$ = 500 nm and 700 nm; third and fourth rows, respectively), to assess whether we can reliably distinguish between Pacific (ocean-dominated) and Atlantic (mixed land–ocean) viewing geometries. The simulated values at single wavelengths in both reflected and polarized light are derived from the reference phase curves presented in \cite{Roccetti2025a}, with the shaded regions indicating the 1$\sigma$ cloud variability spread. }
    \label{fig:phase_curve}
\end{figure*}
In Fig. \ref{fig:phase_curve}, we present the phase curves from both our observations and simulations as a function of different metrics: the spectral slope $\gamma$ in the visible (first row), the B–R broadband filter difference (second row), and the reflectance and polarization at single wavelengths (500 and 700~nm; third and fourth rows, respectively). Comparing the diagnostic power of these approaches helps to identify the most effective observational strategies for retrieving information about different surface viewing geometries of the Earth seen as an exoplanet and to understand the sensitivity of different metrics to the various surface scenarios.\\
\indent Analyzing the spectral slope in the VIS reflectance, we find that the Atlantic (mixed surface) configurations, indicated by red stars, tend to cluster with shallower slopes compared to the Pacific (ocean-dominated) geometries. This trend holds until phase angles exceed ~130$^\circ$, where the reflectance spectra flatten due to the strong contribution of ocean glint. A clear distinction between ocean and mixed surface compositions is evident in the spectral slopes at phase angles smaller than quadrature. Beyond quadrature, the differences become less significant. It is important to note that we simulate 53 specific geometries corresponding to the observations, and Earth does not perfectly represent a pure ocean or land surface planet. In particular, for the Atlantic case, the land–ocean fraction plays a key role in shaping the spectral slope, especially depending on whether the sunglint falls over ocean or land.\\
\indent A similar but opposite trend appears in polarization: Atlantic (mixed) configurations display steeper polarization slopes than Pacific (ocean) ones at comparable phase angles. This difference is more pronounced at phase angles below quadrature than in reflectance. At larger phase angles over the Pacific, the ocean glint substantially enhances polarization, resulting in a flatter slope. In this polarization panel (top right in Fig. \ref{fig:phase_curve}), both the simulations (stars) and Earthshine observations (dots) are included, with observed slopes broadly matching the simulated values. As explained in Section \ref{sec:comparison}, we consistently exclude the epochs from A.1 to C.2 in subsequent analyses, and these epochs are shown in opaque colors.\\
\indent In the second row of Fig. \ref{fig:phase_curve}, we examine the sensitivity of the B-R broadband index to distinguish between Pacific and Atlantic configurations. In reflectance, the B–R value tends to be smaller for Atlantic cases relative to Pacific ones at the same phase angle, though the clustering is less distinct than for the spectral slope. A similar trend is observed in polarization, where Atlantic epochs exhibit larger B–R values than Pacific ones, though again without clear separation between the two groups.\\
\indent The third and fourth rows show the phase curves of reflectance and polarization at single wavelengths: 500 and 700~nm. For these plots, we overlay the set of reference phase curves for an ocean (Pacific) and a mixed (Atlantic) surface, as presented in \cite{Roccetti2025a, Roccetti2025b}, corresponding to the 3D CG EXP-RAN x3 scenario. While Earth never fully resembles a pure ocean planet and its surface composition varies with viewing geometry, these models guide interpretation and highlight existing trends. The shaded regions indicate the 1$\sigma$ spread due to cloud variability, calculated from averaging phase curves with different cloud fields from different days of the year \citep{Roccetti2025a}. Comparing the simulations for each Earthshine epoch (stars) with the model phase curves allows us to assess how representative these models are across the Earthshine catalog.\\
\indent All reflectance simulations fall within the 1$\sigma$ cloud variability range, while some polarization simulations extend beyond it. This suggests that while disk-integrated reflectance is relatively well constrained, polarization is more complex to model, but also richer in diagnostic information about surface, atmospheric, and cloud characteristics.\\
\indent Reflectance phase curves at 500 and 700~nm show similar behavior, with higher values at 500~nm due to Rayleigh scattering. However, the differences between the two model phase curves fall within the cloud variability, making it difficult to distinguish surface properties from reflectance at a single wavelength in the VIS. At 500~nm in polarization, the phase curves for the ocean and mixed cases are also similar and within the cloud variability spread. Observed and simulated Atlantic epochs are slightly below the reference model predictions, occasionally falling outside the 1$\sigma$ range. \\
\indent However, at 700~nm, beyond the Rayleigh scattering regime, we observe a pronounced separation between the ocean and mixed model phase curves, far exceeding the 1$\sigma$ cloud variability. This diagnostic power was previously highlighted in \cite{Roccetti2025a}. Overlaying the individual simulations and observations, we find broad agreement with this trend, reinforcing the value of polarization in distinguishing surface types. In particular, the presence of ocean glint significantly increases polarization when the sunglint is over water, while it contributes little when the glint is hidden over land. Some intermediate epochs fall between the two model curves, representing the scenes not fully described by either a pure ocean or a glint-obscured mixed surface. This behavior is consistent with earlier findings, including preliminary simulations of ocean planets with horizontally inhomogeneous clouds with varying cloud fractions by \cite{Trees2019}, as well as indications of ocean presence in Earthshine polarization observations in the NIR reported by \cite{Takahashi2021}.

\subsection{Cloud properties}

\begin{figure*}
    \centering
    \includegraphics[width=0.9\linewidth]{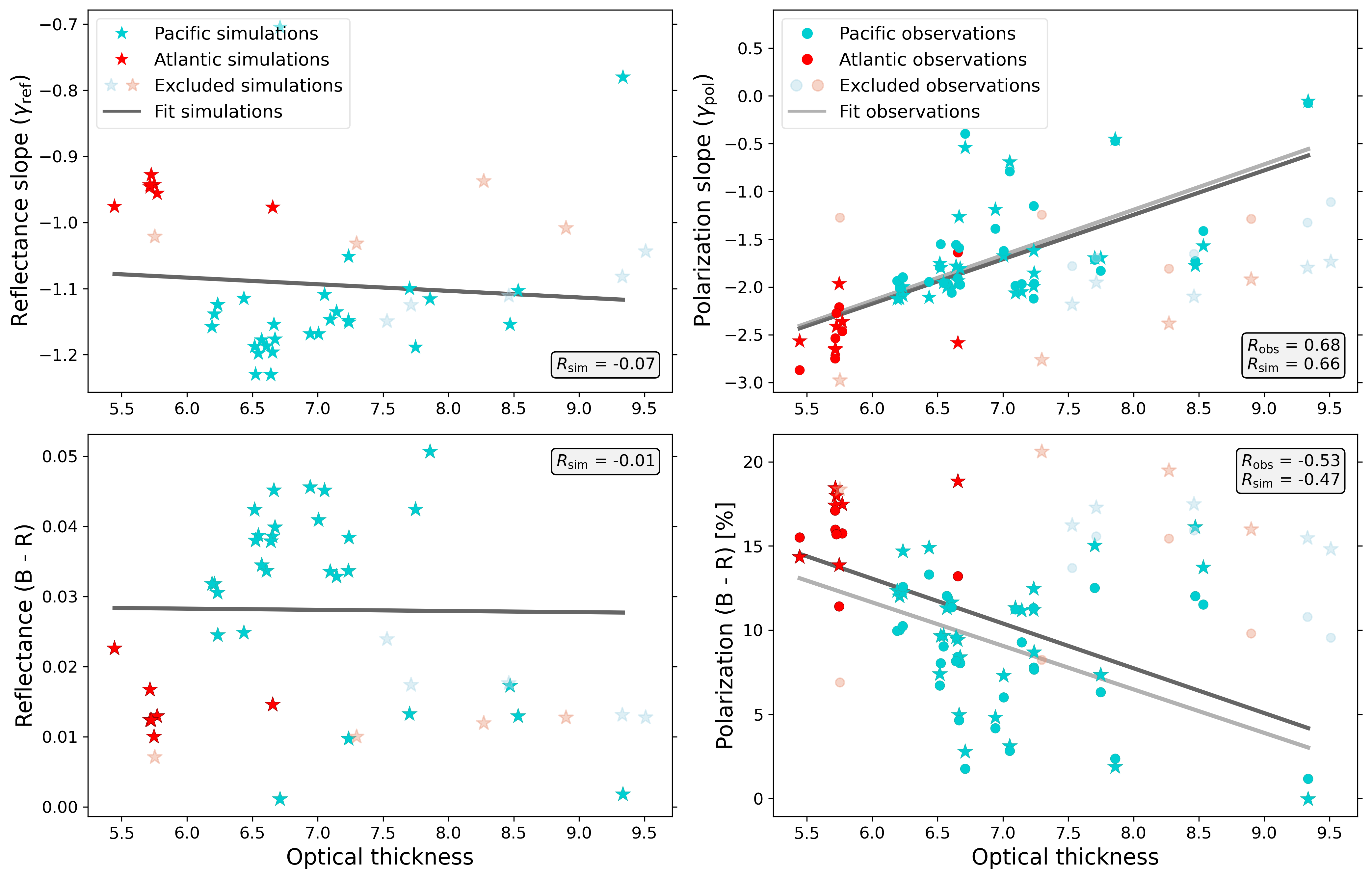}
    \caption{Correlations between the cloud optical thickness and the VIS spectral slopes (first row) and the B-R broadband filter differences (second row) for reflectance simulations (first column) and polarized simulations and observations (second column). We provide the Pearson correlation coefficient ($R$) values and the linear fits for the simulations (black) and observations (gray).}
    \label{fig:tau}
\end{figure*}

\begin{figure*}
    \centering
    \includegraphics[width=0.9\linewidth]{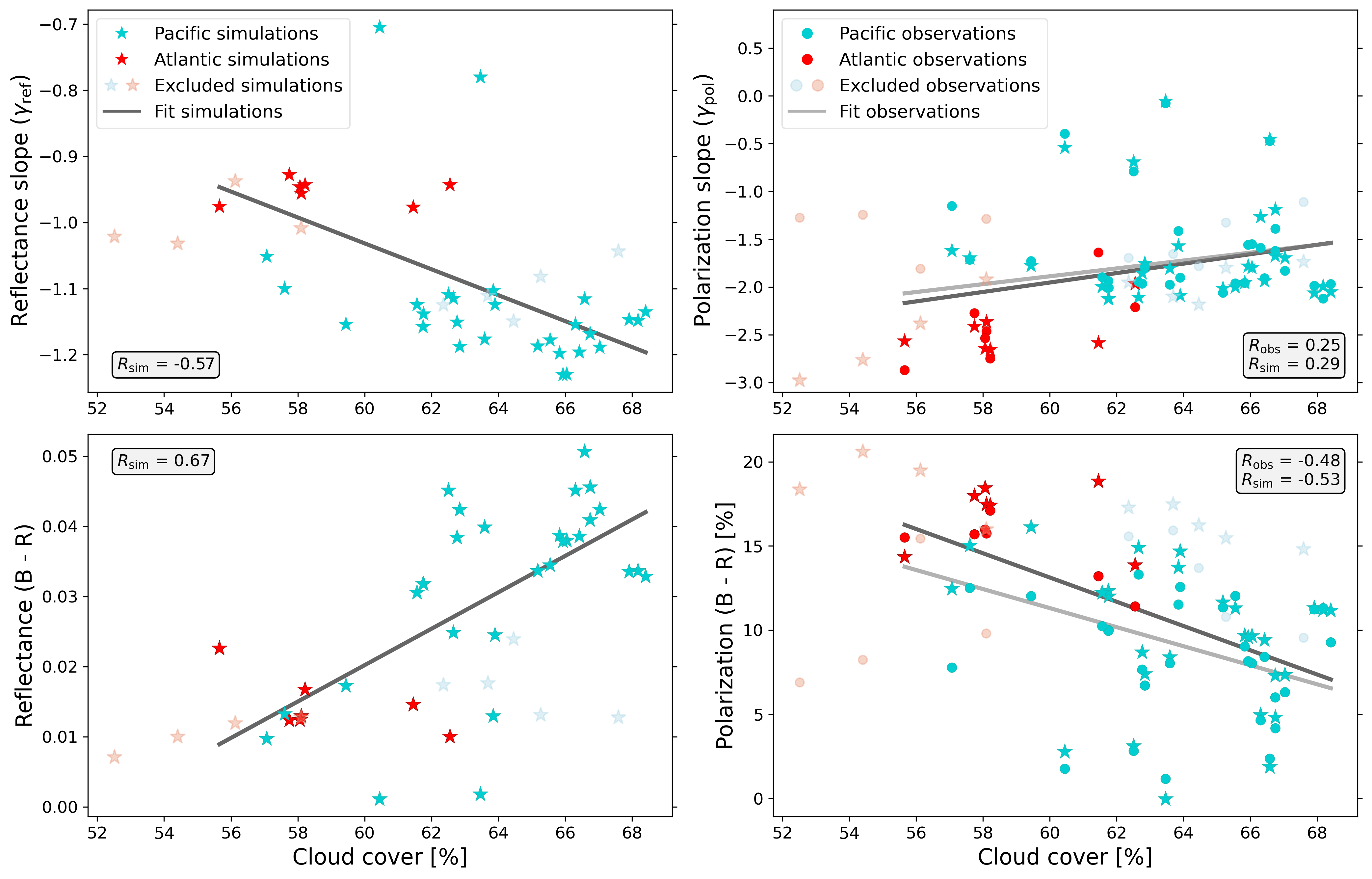}
    \caption{Correlations between the cloud cover and the VIS spectral slopes (first row) and the B-R broadband filter differences (second row) for reflectance simulations (first column) and polarized simulations and observations (second column). We provide the Pearson correlation coefficient ($R$) values and the linear fits for the simulations (black) and observations (gray).}
    \label{fig:cloud_cover}
\end{figure*}

After evaluating the sensitivity to various surface conditions, we now focus on cloud properties. Specifically, we examine the averaged cloud optical thickness $\tau$ and cloud cover ($\text{cc}$) for the different viewing geometries of our Earthshine catalog. The true values of $\tau$ and $\text{cc}$ are derived from the ERA5 reanalysis product for the cloud field used in the simulations (the closest in time), as described in \cite{Roccetti2025a}, and are reported in Table \ref{table}. These values are calculated only for the illuminated side of the planet, as seen from the Moon in the corresponding geometry.\\
\indent In Fig. \ref{fig:tau}, we explore the correlation between cloud optical thickness and both the VIS spectral slope ($\gamma$ from Eq. \ref{eq:fit}) and the B-R broadband differences in both reflectance and polarization. We also tested the single wavelength cases, as in \ref{sec:ocean-land}, but we did not find any significant trends. To evaluate whether any linear correlation exists between the simulations and/or observations, we use the Pearson correlation coefficient (R-value), which is reported in each panel. For reflectance, we report only the R-value for simulations ($R_{\text{sim}}$), while for polarization, the Pearson correlation coefficient is calculated for both simulations and observations ($R_{\text{obs}}$).\\
\indent From the R-values, we conclude that there is no significant correlation between either the reflectance slope or the B-R difference and the cloud optical thickness. However, a moderate linear correlation is observed for the polarization slope, which becomes moderately strong in the case of the B-R difference. This can be understood physically as increasing cloud optical thickness enhances multiple scattering within the cloud layer, which generally reduces the degree of polarization. At the same time, the relative contribution of Rayleigh scattering (which has strong wavelength dependence) decreases compared to Mie scattering (which varies weakly with wavelength in this spectral range), resulting in a flatter polarization slope. This suggests that a steeper VIS slope or a larger broadband B-R difference in polarization corresponds to a smaller cloud optical thickness in the corresponding viewing geometry.  Additionally, we observe a clustering of the Atlantic (mixed) epochs in the lower optical thickness range, consistent with $\tau$ predictions for a mixture of land and ocean sceneries, as previously reported by the statistical analysis of ERA5 cloud fields in \cite{Roccetti2025a}.\\
\indent Next, we examine the same trends, but with cloud cover instead of optical thickness (Fig. \ref{fig:cloud_cover}). Cloud cover values are calculated for the illuminated viewing geometry of each Earthshine epoch and range from 55\% to more than 69\%, as obtained from the ERA5 cloud fields. With cloud cover, we find a weak correlation with the polarization slope. Instead, we see a moderate linear correlation with the polarization B-R differences, with an R-value around -0.5 for both simulations and observations. This negative correlation indicates that a larger cloud cover is associated with a flatter spectrum from the B to the R filter. For cloud cover, we observe larger R-values in reflectance, suggesting that this is a better diagnostic metric for assessing cloud cover, particularly through the B-R coefficient. In contrast, the best diagnostic metric for optical thickness was found to be the VIS spectral slope in polarization. In reflectance, we observe a moderate anticorrelation between the spectral slope and cloud cover, while the B-R difference exhibits a moderately strong linear correlation with an R-value of 0.67.\\
\indent Thus, we find that polarization is more sensitive to the cloud optical thickness, while reflectance is more sensitive to cloud cover. Furthermore, the spectral slope provides more diagnostic information on optical thickness in polarization, whereas the B-R difference contains more sensitivity in reflectance. Moreover, it is important to note that assessing cloud properties is challenging due to the degeneracy of various parameters such as optical thickness, cloud cover, cloud deck height, and cloud droplet size. Thus, these moderately strong linear correlations with optical thickness and cloud cover are highly relevant and may pave the way for novel methods of discriminating cloud properties on exoplanets. Our findings suggest that combining reflected and polarized light could be a promising avenue for such investigations.

\subsection{Vegetation red edge}

\begin{figure*}
    \centering
    \includegraphics[width=0.9\linewidth]{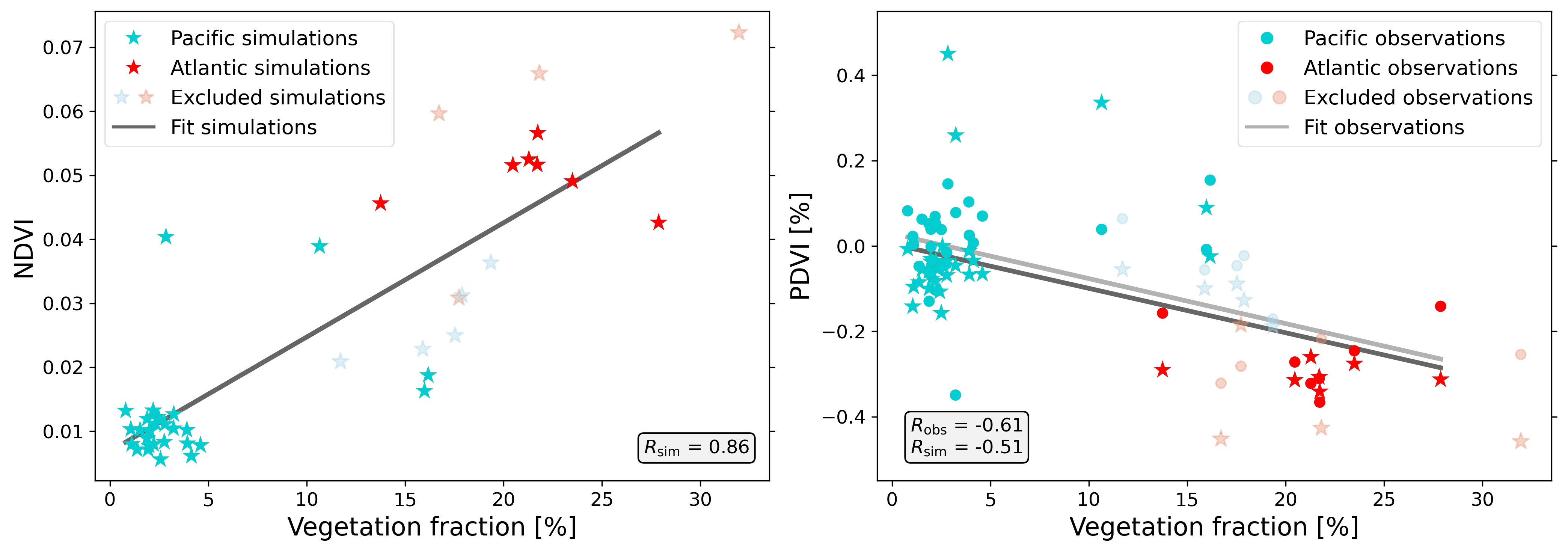}
    \caption{Correlation between the detection of the VRE feature and the vegetation fraction of the viewing geometry. The left panel shows the correlation with the NDVI, while the right panel shows the correlation with the PDVI. The Pearson correlation coefficients ($\text{R}$) and the linear fits for both simulations (black) and observations (gray) are provided.}
    \label{fig:VRE}
\end{figure*}

As a final feature, we examine the detection of the VRE in both the simulated reflectance spectra and the observed and simulated spectra in polarization. The VRE is characterized by an increase in reflectivity (and a decrease in polarization) between 700 and 800~nm due to chlorophyll. In reflectance, the VRE is quantified using the NDVI, defined as
\begin{equation}
    \text{NDVI} = \frac{\text{NIR} - \text{R}}{\text{NIR} + \text{R}},
\end{equation}
where the $\text{NDVI}$ is calculated from the continuum in the NIR range (748–758~nm and 769-778~nm in order to exclude the $O_2$-A band feature) and the red spectral range ($\text{R}$) between 675 and 685~nm, normalized by their sum. In polarization, the VRE is observed as a decrease in the spectral continuum in the same wavelength range. We calculate the ($\text{PDVI}$) as
\begin{equation}
    \text{PDVI} = \text{NIR} - \text{R}.
\end{equation}
Following \cite{Sterzik2019}, we apply a normalization procedure to estimate the $\text{PDVI}$ from both observed and simulated polarization spectra. Specifically, we fit a fourth-order polynomial to the spectra in the 530–890~nm range, excluding regions affected by O$_2$ and H$_2$O absorption features. We then subtract this fit from the original spectra. Unlike the $\text{NDVI}$, the $\text{PDVI}$ is not normalized by the term $\text{NIR} + \text{R}$, in order to avoid unphysical behavior when the denominator approaches zero. The resulting $\text{PDVI}$ values are consistent with those reported in \cite{Sterzik2019}.\\
\indent The fractional vegetation coverage for each viewing geometry is calculated using the MODIS surface type maps. We consider all pixels containing grass or vegetation percentages and sum their vegetation fractions, accounting also for cloudy pixels. The results are reported in Table \ref{table}, along with the percentage of land fraction, derived using a land-sea mask to distinguish between pixels over the ocean and those over land. Additionally, we account for the real dimensions of the pixels and correct for their projected area in our vegetation and land fraction estimations. For each pixel, we first compute its physical area by converting its latitude and longitude span into the area of the corresponding spherical quadrangle on Earth’s surface. We then determine whether the pixel is illuminated by the Sun by calculating the dot product between the unit vector pointing toward the Sun and the surface normal vector of the pixel. If this dot product is negative, the pixel lies in shadow and is considered not illuminated. For the illuminated pixels, we project their area in the direction of the Moon by scaling the surface area by the dot product between the pixel’s normal vector and the unit vector pointing toward the Moon. This projection accounts for viewing geometry, giving greater weight to pixels directly facing the Moon and diminishing the contribution of those observed at a slant angle.\\
\indent In Fig. \ref{fig:VRE}, we show the correlations between the $\text{NDVI}$ (left panel) and $\text{PDVI}$ (right panel) as a function of the vegetation fraction. We clearly observe that the Atlantic (mixed) geometry epochs cluster at larger vegetation fractions, while the Pacific (ocean) points form a compact cluster at very small vegetation fractions. Some Pacific epochs appear in transition between a nearly pure ocean configuration and one with a larger fraction of land and vegetation coverage.\\
\indent Calculating the Pearson correlation coefficients, we find a strong linear correlation between the $\text{NDVI}$ and vegetation fraction, indicating a strong detection of the VRE in the reflectance spectra of Earth as an exoplanet. We obtain a maximum $\text{NDVI}$ of around 0.07, consistent with previous studies such as \cite{Tinetti2006a}. For reference, the $\text{NDVI}$ values for a satellite image over a fully forested area are typically around 0.3, so a value of 0.07 for a vegetation cover fraction of 25\% is in line with theoretical expectations.\\
\indent For the $\text{PDVI}$, we also observe a moderately strong inverse correlation with the vegetation fraction, as expected due to the decrease in polarization slope. Thus, while polarization spectra are sensitive to the VRE, this feature is more prominent in reflectance.\\
\indent Furthermore, we calculate the Pearson coefficients for the $\text{NDVI}$ ($R_{\text{sim}}$ = 0.88) and $\text{PDVI}$ ($R_{\text{obs}}$ = -0.67 and $R_{\text{sim}}$ = -0.63) as a function of land fraction. This plot is not shown in the paper, but it is very similar to the vegetation fraction case (Fig. \ref{fig:VRE}), and it displays slightly stronger correlations with the land fraction. This could be because the land fraction provides a more accurate reflection of the surface composition, whereas the vegetation coverage is based on simplistic assumptions about which pixels should be considered vegetated. Additionally, the MODIS surface type maps represent yearly averages, which may not accurately capture seasonal variations in vegetation, especially over savannas and mixed land cover regions.

\section{Discussion}

Through this catalog of Earthshine polarization observations, spanning phase angles from approximately 37$^\circ$ to 136$^\circ$, we assess several key indicators of Earth’s habitability as it would appear to a distant observer. Polarimetric Earthshine data reveal a wealth of information on both surface and atmospheric properties, crucial components in evaluating planetary habitability.\\
\indent First, using observations focused on the Pacific (ocean-dominated) geometry, \cite{Sterzik2020} showed that the polarization cloudbow signature enables the retrieval of microphysical cloud properties. Specifically, the data allowed us to assess that Earth's clouds are composed of liquid water droplets with an effective radius of approximately 6~$\mu$m. This finding marked the first time such cloud microphysical properties were inferred from disk-integrated spectra, highlighting the unique diagnostic power of polarization.\\
\indent Additionally, in this work, we demonstrate sensitivity to the ocean glint signature at large phase angles in the polarization continuum, evidenced by increased polarization levels in spectra that include specular reflection. The successful detection of ocean glint confirms the presence of surface liquid water and, together with the atmospheric clouds, supports the existence of an active hydrological cycle. Moreover, by analyzing both the spectral and temporal variability in the polarization VIS slope, building on the earlier work by \cite{Sterzik2019}, we assess the coexistence of ocean and land surfaces. This further reinforces the idea that Earth’s polarized signal encodes rich information about surface composition and heterogeneity.\\
\indent When combined with potential detections of atmospheric biosignature gases such as O$_2$, CH$_4$, and H$_2$O (though outside the scope of this paper), spectropolarimetry emerges as a powerful tool for assessing both planetary habitability and possible evidence of an active biosphere on the planet.\\
\indent The diagnostic metrics introduced in this paper must, however, be interpreted in the context of degeneracies that may arise in exoplanet studies. For instance, although the VIS spectral slope shows sensitivity to surface composition, it is also influenced by cloud and aerosol properties \citep[e.g.,][]{Powell2019, Ohno2020}, making its interpretation non-trivial without prior knowledge of atmospheric conditions.\\
\indent Nonetheless, our results reveal several spectral metrics with significant diagnostic power. These insights are essential for shaping the scientific requirements of future telescopes and instruments. For example, we find that broadband filters offer limited sensitivity to surface features, though they remain useful for assessing cloud cover. In contrast, polarization, especially at single wavelengths beyond the Rayleigh scattering regime, proves to be uniquely capable of distinguishing surface geometries, such as ocean-only versus mixed land-ocean configurations. Such distinctions are not achievable through reflectance alone. This work highlights the benefit of a complementary observational strategy that combines reflectance and polarization measurements. This dual approach offers a more comprehensive characterization of Earth-like planets and is likely to be valuable in future studies of habitable exoplanets.

\section{Conclusions}

In this work, we have presented the first effort to simulate a large catalog of 53 Earthshine polarization spectra, spanning a decade of observations from 2011 to 2020. The simulations were conducted using the advanced Monte Carlo radiative transfer model MYSTIC, incorporating a state-of-the-art treatment of 3D cloud properties and wavelength-dependent surface albedo maps. A physically consistent model for ocean specular reflection in both reflectance and polarization was also included \citep{Roccetti2025a}. In parallel, we significantly improved the quality of previously published Earthshine observations \citep{Sterzik2012, Sterzik2019, Sterzik2020}. This was achieved by refining the data reduction and slitlet averaging to reduce contamination from Moonshine and sky background, especially critical for high phase angle observations.\\
\indent With this improved observational dataset, we robustly validated our modeling framework, demonstrating its ability to reproduce the polarization spectra of Earth as seen from afar. This represents a more stringent validation than reflectance comparisons alone, given the enhanced sensitivity of polarization to both surface and atmospheric properties, as further discussed in this paper and in earlier work of our paper series \citep{Roccetti2025a, Roccetti2025b}.\\
\indent Our model successfully reproduces the observed spectral slopes, polarization levels, and most of the absorption features, a substantial achievement considering the complexity involved in modeling Earthshine polarization \citep{Emde2017, Gordon2023}. This success is primarily due to improved cloud representation, particularly through the 3D CG developed in \cite{Roccetti2025a}. Accounting for sub-grid cloud variability and heterogeneity allows us to accurately simulate even the most challenging high phase angle spectra, which are dominated by a complex interplay between ocean glint and overlying 3D cloud structure.\\
\indent We also accurately reproduce the Atlantic (mixed surface) viewing geometry spectra, thanks to the incorporation of HAMSTER wavelength-dependent albedo maps \citep{Roccetti2024}. These maps avoid overestimating the reflectivity of vegetated and soil-covered surfaces \citep{Roccetti2025a}, allowing us to match the continuum of the observations.\\
\indent However, our simulations were unable to reproduce the first eight observed spectra (from A.1 to C.2), which appear significantly flatter than later spectra acquired at similar phase angles and geometries. We did not find a definitive explanation, although possible causes include observational systematics or changes in the global atmospheric state, without clear supporting evidence.\\
\indent We then used both the simulated and observed polarization spectra to assess the diagnostic potential of reflectance and polarization in probing surface viewing geometry, cloud properties, and the VRE. We evaluated the performance of various spectral metrics, including the VIS spectral slope, broadband B--R color differences, and single-wavelength values. Our main findings are:
\begin{itemize}
    \item The VIS spectral slope in both reflectance and polarization effectively separates Pacific (ocean-dominated) and Atlantic (mixed surface) epochs. Atlantic observations show flatter slopes in reflectance and steeper slopes in polarization compared to the Pacific ones. The B--R color index shows limited diagnostic power.
    \item Polarization at wavelengths beyond the Rayleigh scattering regime effectively distinguishes surface types, confirming model predictions \citep{Roccetti2025a}. This result demonstrates the value of polarization for surface characterization in disk-integrated exoplanet spectra.
    \item For the first time, we demonstrate that Earthshine polarization spectra are sensitive to cloud properties. Specifically, the VIS polarization slope shows a moderately strong correlation with cloud optical thickness, while the B--R reflectance index better captures the cloud cover of the planet.
    \item We find a strong correlation between the NDVI and vegetation fraction, and a moderately strong anti-correlation between the PDVI and vegetation fraction. These results confirm the detectability of the VRE in disk-integrated spectra, especially in reflectance.
\end{itemize}
In conclusion, our model is now capable of reproducing the majority of the 53 Earthshine polarization spectra acquired with FORS2 at the VLT. This is a non-trivial accomplishment, considering that polarization signals are strongly influenced by viewing geometry, cloud and aerosol properties, surface composition, and solar illumination angle.\\
\indent These results establish a critical benchmark for future modeling approaches and observational diagnostics for rocky exoplanet characterization. In particular, our advanced treatment of clouds and surfaces, introduced in \cite{Roccetti2025a}, is the ground-truth framework for simulating Earthshine observations and can inform other modeling approaches for studying the Earth as an exoplanet. These results are instrumental in developing retrieval techniques for reflected light observations of rocky exoplanets, especially in the context of next-generation observatories such as the ELT and the mission concept HWO.

\section{Data availability}
\label{sec:data}
The observational data and corresponding simulations are publicly available on GitHub\footnote{\url{https://github.com/giulia-roccetti/Earth_as_an_exoplanet_Part_III}}. The same repository also includes a Jupyter Notebook containing all plotting routines used to reproduce the figures in this work.

\begin{acknowledgements}
This paper is based on observations collected at the European Southern Observatory (ESO) under the P87.C-0040, P90.C-0096, and P104.C-0048 programmes. GR and JVS were supported by the Munich Institute for Astro-, Particle and BioPhysics (MIAPbP), which is funded by the Deutsche Forschungsgemeinschaft (DFG, German Research Foundation) under Germany´s Excellence Strategy – EXC-2094 – 390783311. 
\end{acknowledgements}

\bibliographystyle{aa} 
\bibliography{bibliography} 
\clearpage
\onecolumn
\begin{appendix}
\section{Summary table of observed and simulated Earthshine spectra}
\label{table}
\begin{center}
\begin{adjustbox}{angle=90}
\begin{minipage}{\textheight}
\centering
\captionof{table}{Overview of Earthshine observational epochs with dates, cloud and surface characteristics, and fitted parameters from observations and simulations.}
\resizebox{0.95\textwidth}{!}{
\begin{tabular}{ccccccccc|ccccc|ccccc|ccccc}
\toprule
& & & & & & & & & \multicolumn{5}{c|}{Observations} & \multicolumn{10}{c}{Simulations} \\ 
ID & date & $\alpha$ [°] & Geom. & Grism & $\tau$ & cc [\%] & Land [\%] & Veg. [\%] & $\gamma_{\text{pol}}$ & $(B-R)_{\text{pol}}$ [\%] & $P_{500}$ [\%] & $P_{700}$ [\%] & PDVI & $\gamma_{\text{ref}}$ & $(B- R)_{\text{ref}}$ & $R_{500}$ & $R_{700}$ & NDVI & $\gamma_{\text{pol}}$ &  $(B-R)_{\text{pol}}$ [\%] & $P_{500}$ [\%] & $P_{700}$ [\%] & PDVI \\
\midrule
A.1 & 2011-04-24 & 99 & A & 300V & 7.30 & 54.41 & 33.07 & 21.82 & -1.08 & 8.23 & 22.87 & 16.82 & -0.22 & -1.03 & 0.01 & 0.05 & 0.05 & 0.07 & -2.76 & 20.60 & 25.54 & 9.84 & -0.43 \\
A.2 & 2011-04-24 & 98 & A & 300V & 8.90 & 58.10 & 20.86 & 17.74 & -1.29 & 9.79 & 22.47 & 14.80 & -0.28 & -1.01 & 0.01 & 0.06 & 0.06 & 0.03 & -1.92 & 15.98 & 24.98 & 13.13 & -0.18 \\
A.3 & 2011-04-25 & 87 & A & 300V & 8.27 & 56.13 & 25.16 & 16.72 & -1.81 & 15.42 & 26.79 & 14.83 & -0.31 & -0.94 & 0.01 & 0.07 & 0.07 & 0.06 & -2.38 & 19.48 & 26.34 & 11.73 & -0.45 \\
B.1 & 2011-06-08 & 76 & P & 300V & 7.53 & 64.46 & 17.15 & 15.89 & -1.71 & 13.68 & 25.18 & 16.11 & -0.06 & -1.15 & 0.02 & 0.09 & 0.08 & 0.02 & -2.18 & 16.23 & 23.12 & 11.06 & -0.10 \\
B.2 & 2011-06-08 & 88 & P & 300V & 7.71 & 62.36 & 20.93 & 19.35 & -1.64 & 15.56 & 29.96 & 19.20 & -0.17 & -1.12 & 0.02 & 0.07 & 0.06 & 0.04 & -1.95 & 17.28 & 26.80 & 13.81 & -0.19 \\
B.3 & 2011-06-09 & 89 & P & 300V & 8.46 & 63.69 & 18.81 & 17.53 & -1.62 & 15.92 & 31.25 & 20.10 & -0.05 & -1.11 & 0.02 & 0.07 & 0.06 & 0.02 & -2.10 & 17.48 & 25.69 & 12.65 & -0.09 \\
B.4 & 2011-06-10 & 102 & P & 300V & 9.33 & 65.27 & 19.01 & 17.89 & -1.27 & 10.78 & 27.03 & 19.80 & -0.02 & -1.08 & 0.01 & 0.06 & 0.05 & 0.03 & -1.80 & 15.47 & 25.96 & 14.11 & -0.13 \\
B.5 & 2011-06-10 & 103 & P & 300V & 9.51 & 67.59 & 12.44 & 11.71 & -1.05 & 9.54 & 27.34 & 20.79 & 0.06 & -1.04 & 0.01 & 0.06 & 0.05 & 0.02 & -1.74 & 14.81 & 25.43 & 14.23 & -0.05 \\
C.1 & 2012-10-06 & 112 & A & 300V & 5.75 & 52.52 & 49.97 & 31.96 & -1.10 & 6.96 & 20.03 & 14.17 & -0.30 & -1.02 & 0.01 & 0.04 & 0.04 & 0.07 & -2.98 & 18.35 & 21.92 & 7.80 & -0.46 \\
C.2 & 2012-10-06 & 111 & A & 600I & 6.97 & 57.53 & 27.03 & 16.83 & - & - & - & 14.24 & - & -1.03 & 0.01 & 0.04 & 0.04 & 0.05 & -2.85 & 17.22 & 21.72 & 8.59 & -0.29 \\
D.1 & 2012-12-07 & 81 & A & 300V & 6.66 & 61.46 & 33.80 & 21.74 & -1.64 & 13.20 & 22.52 & 11.10 & -0.37 & -0.98 & 0.01 & 0.08 & 0.08 & 0.06 & -2.59 & 18.84 & 24.18 & 10.04 & -0.34 \\
E.1 & 2012-12-17 & 50 & P & 300V & 6.52 & 62.84 & 3.62 & 1.94 & -1.81 & 6.74 & 12.55 & 7.59 & -0.06 & -1.19 & 0.04 & 0.15 & 0.13 & 0.01 & -1.75 & 7.39 & 12.33 & 6.96 & -0.06 \\
E.2 & 2012-12-18 & 63 & P & 300V & 6.61 & 65.18 & 3.82 & 2.57 & -2.08 & 11.62 & 17.73 & 8.47 & -0.05 & -1.19 & 0.03 & 0.12 & 0.10 & 0.01 & -2.02 & 11.65 & 17.51 & 8.94 & -0.00 \\
E.3 & 2012-12-18 & 63 & P & 600I & 5.96 & 64.56 & 1.90 & 0.83 & - & - & - & 8.42 & - & -1.17 & 0.03 & 0.12 & 0.10 & 0.01 & -1.76 & 11.55 & 17.36 & 8.88 & 0.00 \\
E.4 & 2012-12-19 & 75 & P & 300V & 6.44 & 62.66 & 2.91 & 1.98 & -1.96 & 13.30 & 21.23 & 10.49 & 0.04 & -1.11 & 0.02 & 0.10 & 0.08 & 0.01 & -2.11 & 14.89 & 21.73 & 10.73 & -0.03 \\
E.5 & 2012-12-19 & 76 & P & 600I & 6.32 & 63.57 & 1.07 & 0.40 & - & - & - & 9.82 & - & -1.13 & 0.02 & 0.10 & 0.08 & 0.01 & -2.06 & 14.74 & 21.25 & 10.40 & -0.06 \\
E.6 & 2012-12-19 & 76 & P & 300V & 6.24 & 63.90 & 1.47 & 0.80 & -1.92 & 12.56 & 20.69 & 10.69 & 0.08 & -1.12 & 0.02 & 0.09 & 0.08 & 0.01 & -2.09 & 14.68 & 21.41 & 10.67 & -0.01 \\
F.1 & 2013-02-02 & 107 & A & 300V & 5.75 & 62.55 & 42.18 & 27.88 & -1.82 & 11.40 & 20.17 & 12.16 & -0.14 & -0.94 & 0.01 & 0.06 & 0.05 & 0.04 & -1.97 & 13.86 & 21.41 & 11.05 & -0.31 \\
F.2 & 2013-02-03 & 94 & A & 300V & 5.77 & 58.11 & 38.16 & 23.50 & -2.49 & 15.74 & 21.68 & 10.23 & -0.25 & -0.96 & 0.01 & 0.07 & 0.07 & 0.05 & -2.37 & 17.47 & 23.52 & 10.63 & -0.28 \\
F.3 & 2013-02-03 & 93 & A & 300V & 5.72 & 58.06 & 34.71 & 20.47 & -2.55 & 15.96 & 21.48 & 9.93 & -0.27 & -0.95 & 0.01 & 0.07 & 0.07 & 0.05 & -2.64 & 18.44 & 22.90 & 9.30 & -0.31 \\
F.4 & 2013-02-03 & 92 & A & 300V & 5.73 & 57.75 & 24.56 & 13.75 & -2.28 & 15.69 & 22.87 & 11.58 & -0.16 & -0.93 & 0.01 & 0.07 & 0.07 & 0.05 & -2.41 & 17.98 & 23.93 & 10.54 & -0.29 \\
F.5 & 2013-02-04 & 80 & A & 300V & 5.72 & 58.22 & 36.39 & 21.71 & -2.78 & 17.10 & 20.54 & 7.45 & -0.31 & -0.94 & 0.02 & 0.09 & 0.08 & 0.05 & -2.66 & 17.41 & 21.56 & 8.74 & -0.31 \\
F.6 & 2013-02-04 & 80 & A & 600I & 5.61 & 58.05 & 30.71 & 17.86 & - & - & - & 7.69 & - & -0.95 & 0.02 & 0.09 & 0.09 & 0.05 & -2.69 & 17.47 & 21.42 & 8.53 & -0.34 \\
F.7 & 2013-02-05 & 67 & A & 300V & 5.45 & 55.66 & 35.74 & 21.29 & -2.93 & 15.93 & 18.88 & 6.71 & -0.33 & -0.98 & 0.02 & 0.11 & 0.10 & 0.05 & -2.57 & 14.34 & 18.08 & 7.57 & -0.26 \\
F.8 & 2013-02-05 & 67 & A & 600I & 5.62 & 55.62 & 31.39 & 18.41 & - & - & - & 7.18 & - & -0.98 & 0.02 & 0.11 & 0.10 & 0.05 & -2.65 & 14.75 & 18.07 & 7.18 & -0.29 \\
G.1 & 2013-02-18 & 92 & P & 300V & 8.47 & 59.44 & 3.41 & 3.23 & -1.48 & 12.01 & 25.06 & 15.17 & -0.35 & -1.15 & 0.02 & 0.07 & 0.06 & 0.01 & -1.78 & 16.12 & 27.12 & 14.95 & -0.05 \\
G.2 & 2013-02-19 & 102 & P & 300V & 7.70 & 57.61 & 16.89 & 16.17 & -1.36 & 12.50 & 28.00 & 17.39 & 0.15 & -1.10 & 0.01 & 0.06 & 0.05 & 0.02 & -1.70 & 15.02 & 26.29 & 14.89 & -0.02 \\
G.3 & 2013-02-19 & 102 & P & 600I & 7.29 & 58.66 & 8.55 & 7.87 & - & - & - & 17.82 & - & -1.11 & 0.01 & 0.06 & 0.05 & 0.02 & -1.81 & 14.55 & 25.67 & 14.60 & 0.10 \\
G.4 & 2013-02-19 & 103 & P & 300V & 8.53 & 63.85 & 3.41 & 3.24 & -1.41 & 11.51 & 26.31 & 17.07 & 0.08 & -1.10 & 0.01 & 0.06 & 0.05 & 0.01 & -1.57 & 13.72 & 25.62 & 15.17 & 0.26 \\
G.5 & 2013-02-20 & 113 & P & 300V & 7.24 & 57.07 & 16.79 & 15.98 & -1.18 & 7.77 & 21.40 & 14.53 & -0.01 & -1.05 & 0.01 & 0.05 & 0.04 & 0.02 & -1.62 & 12.45 & 22.76 & 13.22 & 0.09 \\
G.6 & 2013-02-20 & 114 & P & 600I & 8.06 & 59.64 & 4.80 & 4.48 & - & - & - & 12.49 & - & -1.08 & 0.01 & 0.05 & 0.04 & 0.02 & -1.78 & 11.04 & 22.88 & 14.25 & 0.15 \\
G.7 & 2013-02-22 & 135 & P & 300V & 6.71 & 60.45 & 11.74 & 10.65 & -0.39 & 1.76 & 13.88 & 12.77 & 0.04 & -0.70 & 0.00 & 0.03 & 0.03 & 0.04 & -0.54 & 2.77 & 14.66 & 11.92 & 0.34 \\
G.8 & 2013-02-22 & 135 & P & 600I & 8.34 & 62.81 & 3.60 & 3.25 & - & - & - & 11.67 & - & -0.76 & 0.00 & 0.03 & 0.03 & 0.04 & -2.04 & 2.82 & 14.50 & 11.55 & 0.28 \\
G.9 & 2013-02-22 & 136 & P & 300V & 9.34 & 63.47 & 2.96 & 2.84 & -0.08 & 1.16 & 16.49 & 15.72 & 0.15 & -0.78 & 0.00 & 0.03 & 0.03 & 0.04 & -0.06 & -0.04 & 16.98 & 16.01 & 0.45 \\
H.1 & 2019-10-30 & 37 & P & 300V & 7.86 & 66.58 & 7.15 & 4.14 & -0.49 & 2.36 & 12.60 & 10.66 & 0.01 & -1.12 & 0.05 & 0.19 & 0.16 & 0.01 & -0.45 & 1.87 & 12.49 & 10.61 & -0.03 \\
H.2 & 2019-10-31 & 50 & P & 300V & 7.75 & 67.04 & 7.37 & 3.93 & -1.83 & 6.31 & 10.88 & 6.21 & 0.03 & -1.19 & 0.04 & 0.15 & 0.13 & 0.01 & -1.70 & 7.34 & 12.44 & 7.19 & -0.07 \\
H.3 & 2019-11-01 & 50 & P & 300V & 7.01 & 66.75 & 6.29 & 2.76 & -1.60 & 6.00 & 11.53 & 7.15 & -0.04 & -1.17 & 0.04 & 0.15 & 0.13 & 0.01 & -1.68 & 7.29 & 12.47 & 7.22 & -0.04 \\
H.4 & 2019-11-01 & 62 & P & 300V & 6.57 & 65.55 & 7.76 & 3.91 & -1.91 & 12.02 & 19.40 & 10.46 & 0.10 & -1.18 & 0.03 & 0.12 & 0.10 & 0.01 & -2.00 & 11.30 & 17.06 & 8.77 & -0.01 \\
H.5 & 2019-11-02 & 62 & P & 300V & 7.24 & 68.18 & 5.94 & 2.03 & -1.89 & 11.30 & 18.47 & 9.91 & -0.03 & -1.15 & 0.03 & 0.12 & 0.10 & 0.01 & -1.99 & 11.20 & 16.88 & 8.70 & -0.08 \\
H.6 & 2019-11-02 & 62 & P & 300V & 7.10 & 67.91 & 5.46 & 1.53 & -1.96 & 11.23 & 17.86 & 9.32 & 0.06 & -1.15 & 0.03 & 0.12 & 0.11 & 0.01 & -2.06 & 11.31 & 16.72 & 8.43 & -0.05 \\
H.7 & 2019-11-02 & 63 & P & 300V & 7.14 & 68.41 & 6.51 & 2.43 & -1.97 & 9.27 & 14.78 & 7.68 & -0.05 & -1.14 & 0.03 & 0.12 & 0.11 & 0.01 & -2.05 & 11.16 & 16.31 & 8.27 & -0.11 \\
I.1 & 2019-11-30 & 42 & P & 300V & 7.05 & 62.51 & 6.35 & 2.78 & -0.83 & 2.98 & 10.47 & 8.15 & -0.03 & -1.11 & 0.05 & 0.17 & 0.15 & 0.01 & -0.69 & 3.11 & 10.82 & 8.78 & -0.07 \\
I.2 & 2019-12-01 & 53 & P & 300V & 6.67 & 63.59 & 5.51 & 1.99 & -2.01 & 8.23 & 12.97 & 6.57 & 0.00 & -1.18 & 0.04 & 0.14 & 0.12 & 0.01 & -1.81 & 8.39 & 13.60 & 7.51 & -0.04 \\
I.3 & 2019-12-01 & 54 & P & 300V & 7.24 & 62.77 & 5.41 & 1.88 & -2.02 & 7.82 & 12.33 & 6.24 & 0.06 & -1.15 & 0.04 & 0.14 & 0.12 & 0.01 & -1.86 & 8.68 & 14.00 & 7.58 & -0.06 \\
I.4 & 2019-12-02 & 65 & P & 300V & 6.21 & 61.76 & 4.45 & 1.10 & -2.07 & 10.20 & 15.64 & 7.59 & 0.02 & -1.14 & 0.03 & 0.12 & 0.10 & 0.01 & -2.12 & 12.01 & 17.13 & 8.47 & -0.10 \\
I.5 & 2019-12-02 & 65 & P & 300V & 6.19 & 61.75 & 4.38 & 1.06 & -1.97 & 10.06 & 15.95 & 8.14 & 0.03 & -1.16 & 0.03 & 0.12 & 0.10 & 0.01 & -2.13 & 12.33 & 17.75 & 8.75 & -0.14 \\
I.6 & 2019-12-02 & 65 & P & 300V & 6.23 & 61.57 & 5.65 & 2.19 & -1.93 & 10.35 & 16.66 & 8.47 & 0.06 & -1.12 & 0.03 & 0.12 & 0.10 & 0.01 & -2.00 & 12.22 & 18.51 & 9.52 & -0.08 \\
J.2 & 2019-12-30 & 44 & P & 300V & 6.94 & 66.75 & 5.14 & 2.51 & -1.41 & 4.34 & 9.25 & 5.94 & 0.04 & -1.17 & 0.05 & 0.16 & 0.14 & 0.01 & -1.19 & 4.81 & 10.40 & 7.25 & -0.16 \\
J.3 & 2019-12-30 & 44 & P & 300V & 6.67 & 66.31 & 4.93 & 2.31 & -1.61 & 4.82 & 9.69 & 6.04 & -0.12 & -1.15 & 0.05 & 0.16 & 0.14 & 0.01 & -1.27 & 4.96 & 10.39 & 7.02 & -0.05 \\
J.4 & 2019-12-31 & 55 & P & 300V & 6.65 & 66.42 & 3.50 & 1.38 & -1.94 & 8.56 & 14.03 & 7.25 & -0.04 & -1.20 & 0.04 & 0.13 & 0.11 & 0.01 & -1.94 & 9.41 & 14.67 & 7.70 & -0.09 \\
J.5 & 2019-12-31 & 56 & P & 300V & 6.55 & 65.83 & 3.95 & 1.89 & -1.96 & 9.32 & 15.36 & 8.03 & -0.16 & -1.20 & 0.04 & 0.13 & 0.11 & 0.01 & -1.95 & 9.67 & 15.00 & 7.85 & -0.10 \\
K.2 & 2020-01-30 & 56 & P & 300V & 6.52 & 66.05 & 5.48 & 4.60 & -1.56 & 8.14 & 16.07 & 9.45 & 0.06 & -1.23 & 0.04 & 0.13 & 0.11 & 0.01 & -1.80 & 9.65 & 15.94 & 8.77 & -0.07 \\
K.3 & 2020-01-30 & 57 & P & 300V & 6.64 & 65.93 & 3.01 & 2.22 & -1.56 & 8.24 & 16.12 & 9.47 & 0.06 & -1.23 & 0.04 & 0.13 & 0.11 & 0.01 & -1.78 & 9.57 & 15.91 & 8.77 & -0.03 \\
\bottomrule
\end{tabular}
}
\end{minipage}
\end{adjustbox}
\end{center}

\section{Catalog of observations and simulations}
\label{appendix-B}
\begin{figure*}[h!]
    \centering
    \includegraphics[width=0.80\linewidth]{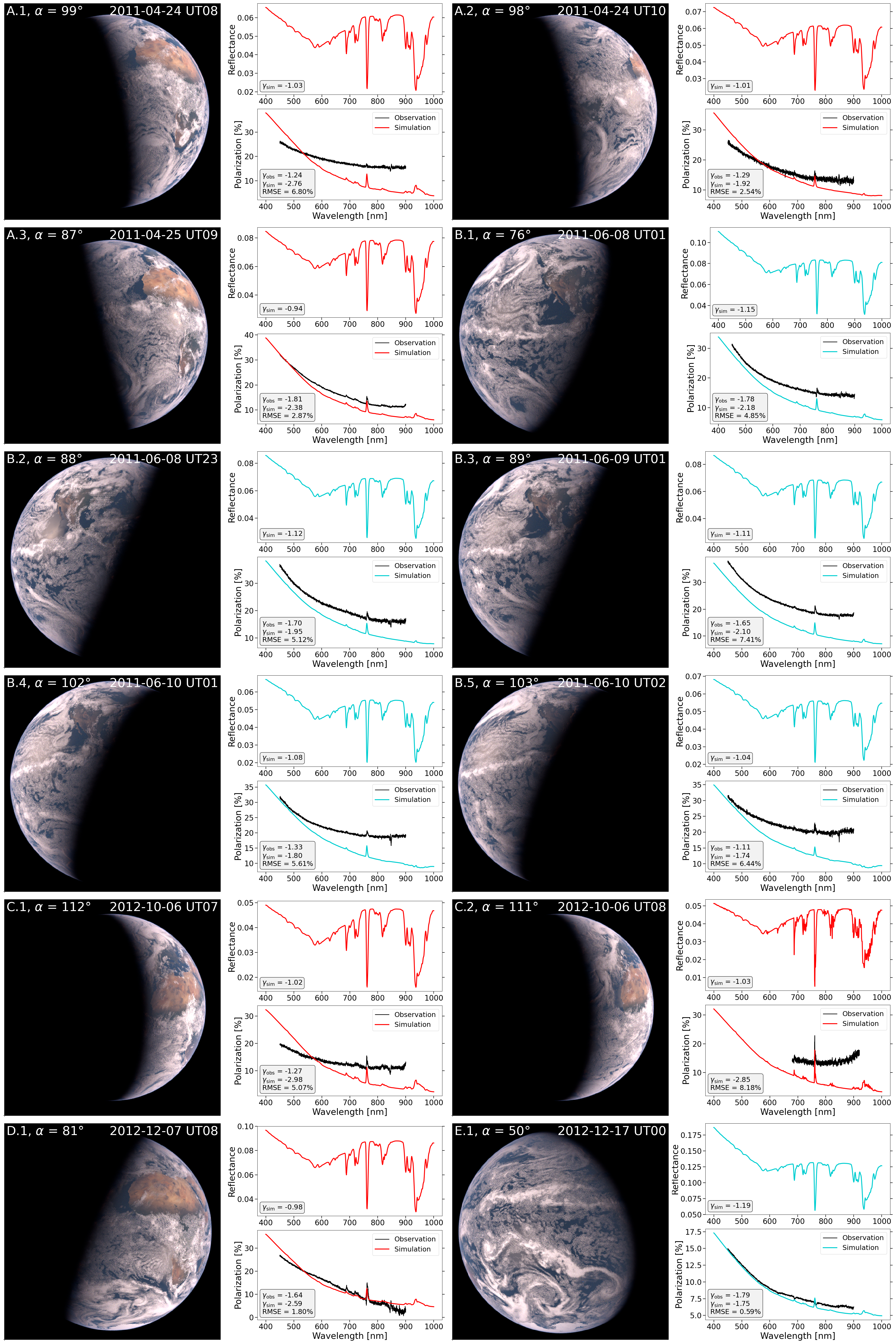}
    \caption{Catalog of observations and simulations from epochs A.1 to E.1.}
    \label{fig:collage1}
\end{figure*}

\begin{figure*}[h!]
    \centering
    \includegraphics[width=0.80\linewidth]{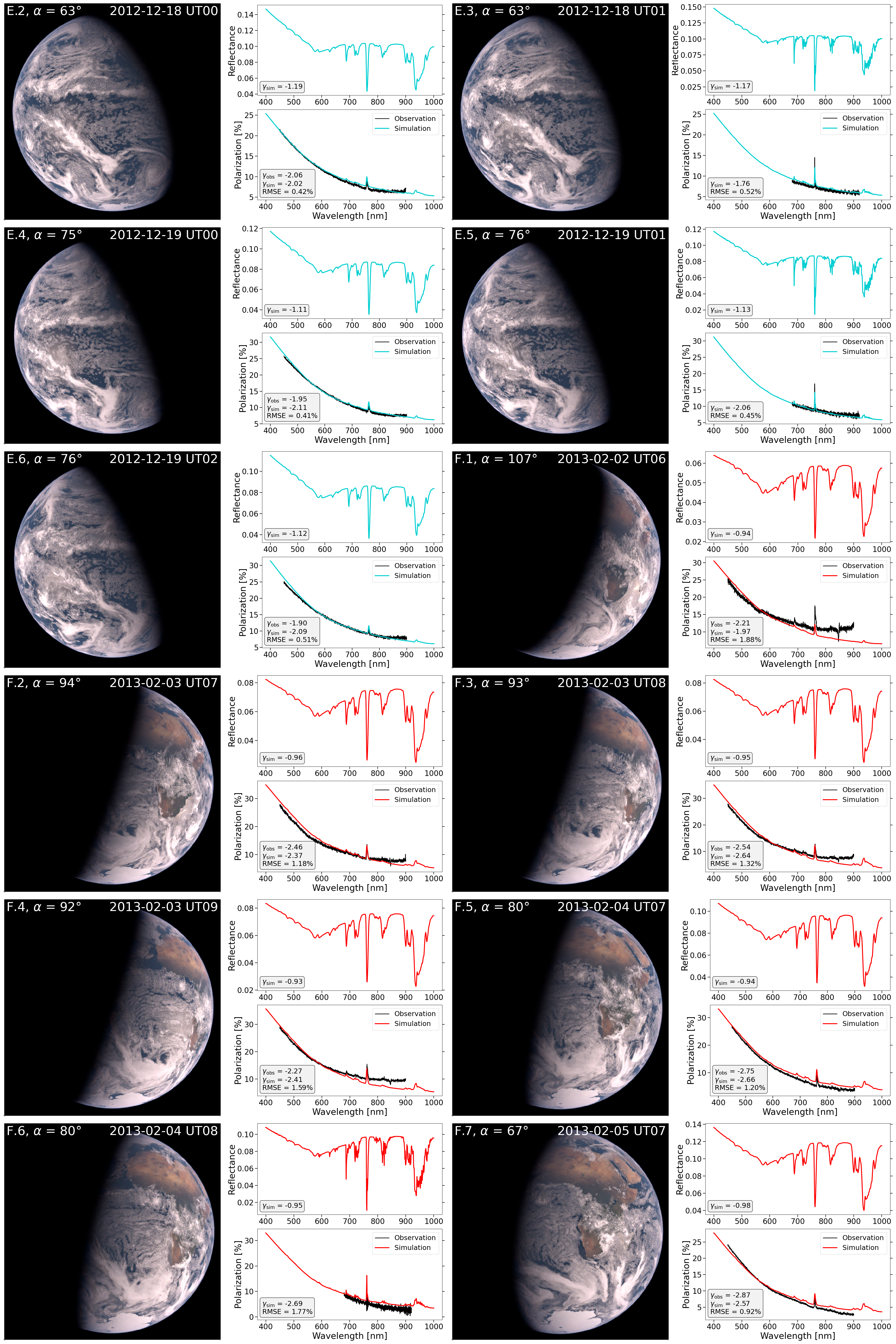}
    \caption{Catalog of observations and simulations from epochs E.2 to F.7.}
    \label{fig:collage2}
\end{figure*}

\begin{figure*}[h!]
    \centering
    \includegraphics[width=0.80\linewidth]{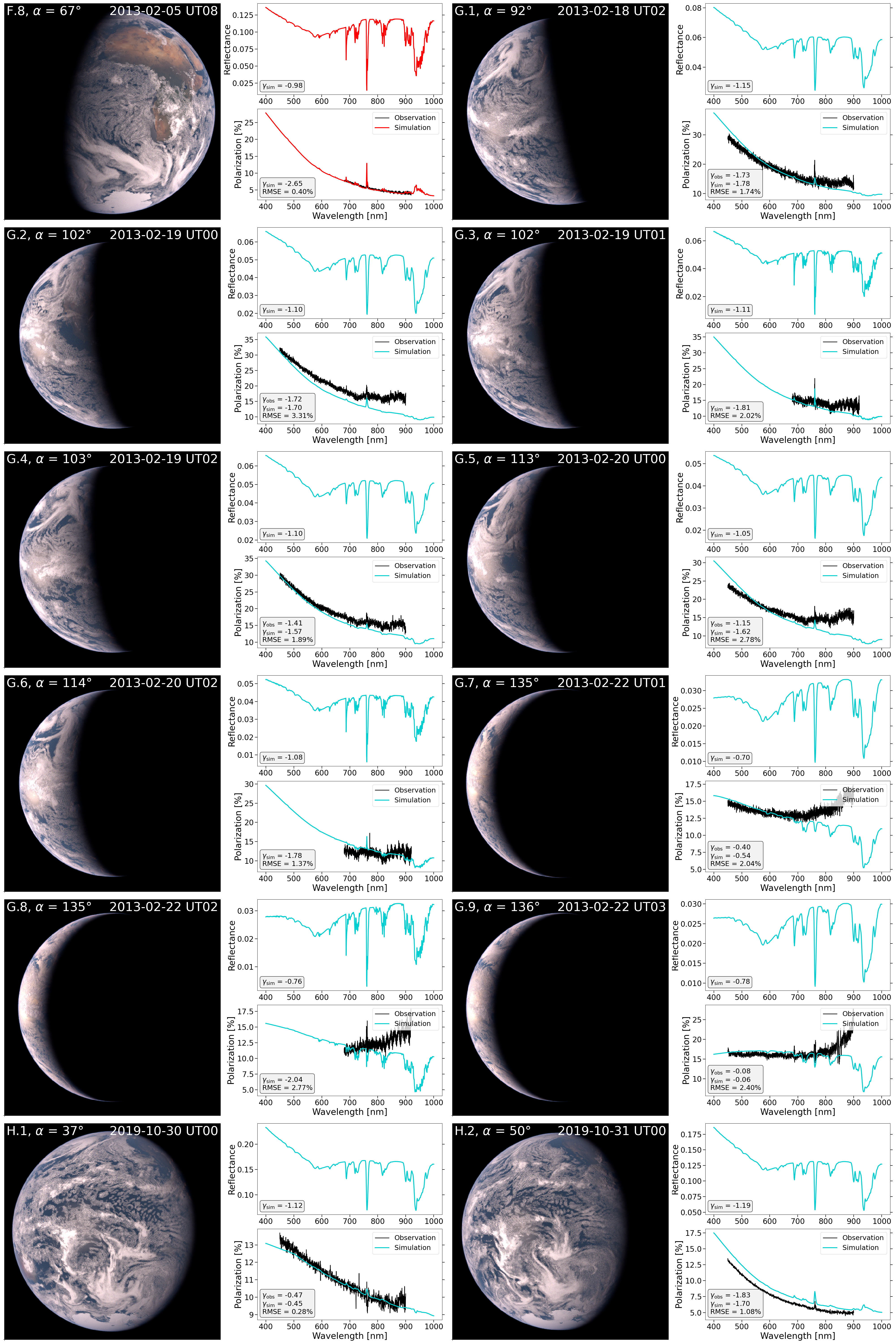}
    \caption{Catalog of observations and simulations from epochs F.8 to H.2.}
    \label{fig:collage3}
\end{figure*}

\begin{figure*}[h!]
    \centering
    \includegraphics[width=0.80\linewidth]{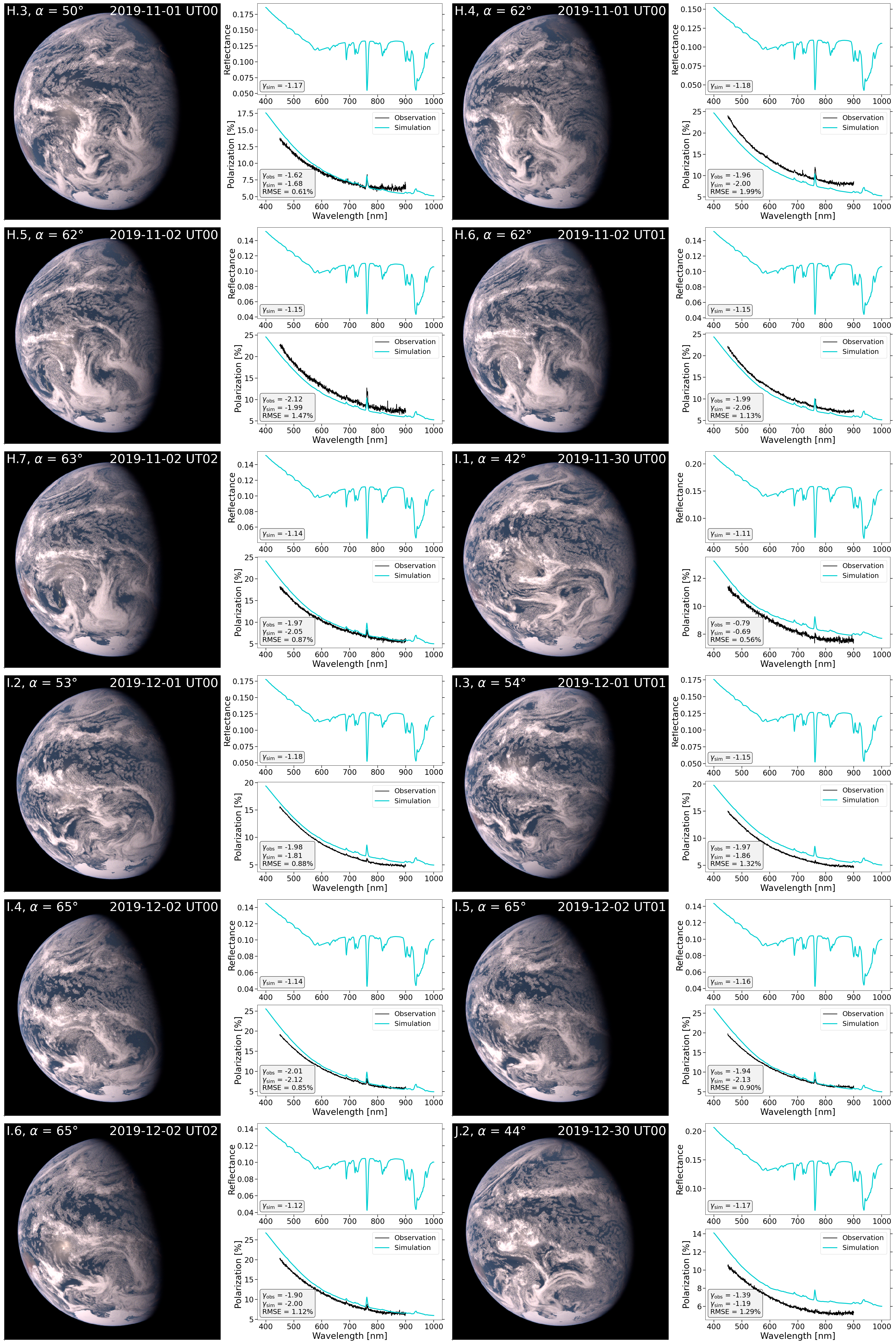}
    \caption{Catalog of observations and simulations from epochs H.3 to J.2.}
    \label{fig:collage4}
\end{figure*}

\begin{figure*}[h!]
    \centering
    \includegraphics[width=0.80\linewidth]{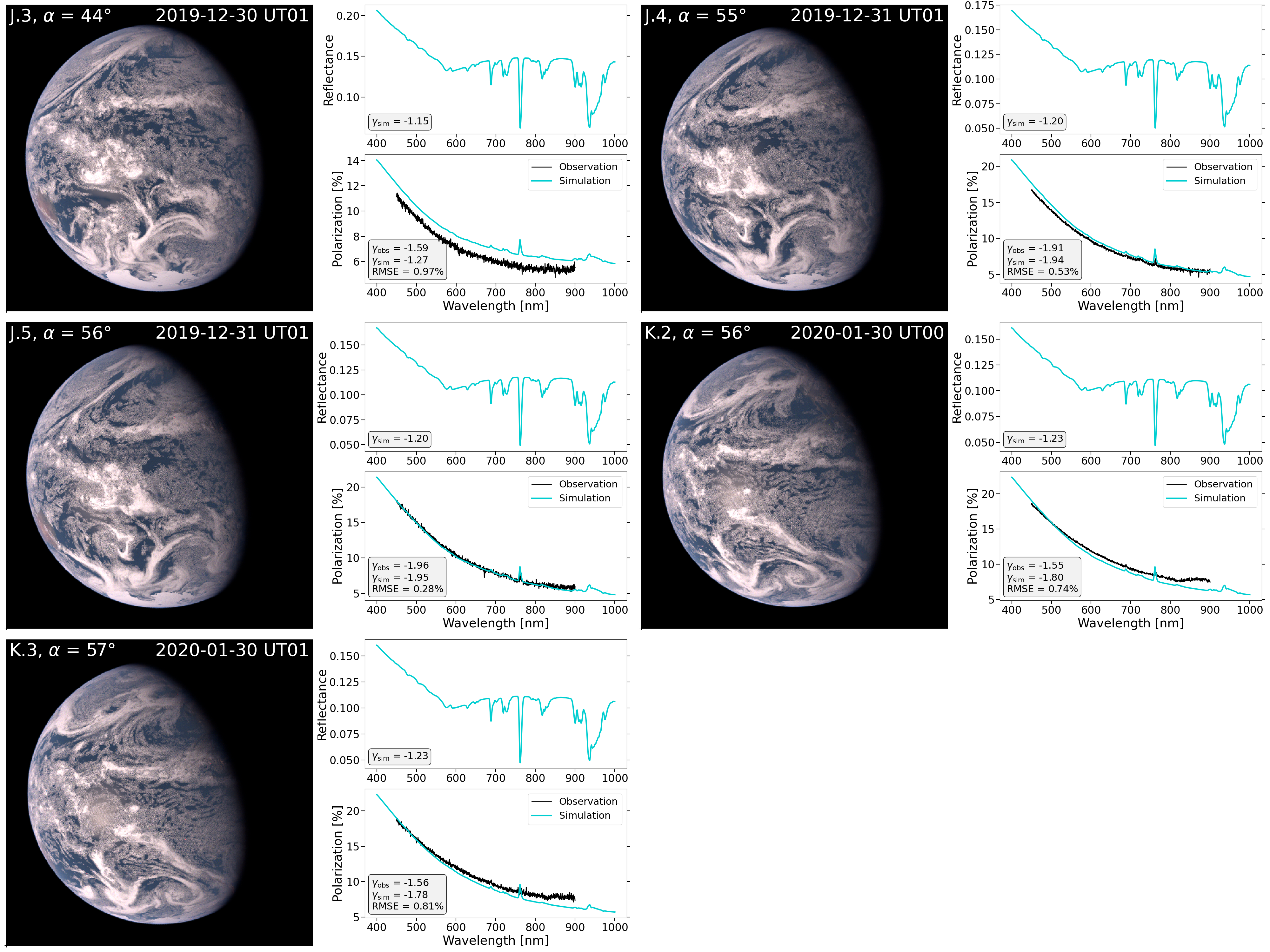}
    \caption{Catalog of observations and simulations from epochs J.3 to K.3.}
    \label{fig:collage5}
\end{figure*}

\end{appendix}

\end{document}